\pdfoutput=1

\documentclass[11pt]{article}

\usepackage{EACL2023}

\usepackage{times}
\usepackage{latexsym}

\usepackage[T1]{fontenc}

\usepackage[utf8]{inputenc}

\usepackage{microtype}

\usepackage{inconsolata}

\usepackage{graphicx}

\usepackage{algorithm}
\usepackage{algpseudocode}
\usepackage{multirow}
\usepackage{xcolor}

\usepackage{booktabs}
\usepackage{amsmath}

\usepackage[strict]{changepage}
\definecolor{darkblue}{rgb}{0.0, 0.0, 0.55}
\usepackage{framed}

\usepackage{xcolor} 
\usepackage[framemethod=TikZ]{mdframed} 
\usepackage{changepage} 

\definecolor{formalshade}{rgb}{255,255,255}
\mdfdefinestyle{MyFrame}{%
    linewidth=0.5pt,
    linecolor=black,
    backgroundcolor=formalshade,
    innertopmargin=-10pt,
    innerleftmargin=2pt,
    innerrightmargin=-3pt,
    innerbottommargin=3pt
}

\newenvironment{formal}{%
  \begin{mdframed}[style=MyFrame]
  \noindent\hspace{-4.55pt}
  \begin{adjustwidth}{}{7pt}%
  \vspace{4pt}\vspace{5pt}%
}{%
  \vspace{1pt}\vspace{1pt}
  \end{adjustwidth}%
  \end{mdframed}%
}

\title{UP5: Unbiased Foundation Model for Fairness-aware Recommendation}%

\title{
\vspace*{-0.5in}
{{\small \hfill EACL 2024}\\
\vspace*{.25in}}
UP5: Unbiased Foundation Model for Fairness-aware Recommendation}

\author{Wenyue Hua, Yingqiang Ge, Shuyuan Xu, Jianchao Ji, Yongfeng Zhang\\
  Department of Computer Science, Rutgers University, New Brunswick, NJ 08854 \\
  \texttt{wenyue.hua,yingqiang.ge,shuyuan.xu,jianchao.ji,yongfeng.zhang}@rutgers.edu }

\begin{document}
\maketitle
\begin{abstract}
Recent advances in Foundation Models such as Large Language Models (LLMs) have propelled them to the forefront of Recommender Systems (RS). Despite their utility, there is a growing concern that LLMs might inadvertently perpetuate societal stereotypes, resulting in unfair recommendations. Since fairness is critical for RS as many users take it for decision-making and demand fulfillment, this paper focuses on user-side fairness for LLM-based recommendation where the users may require a recommender system to be fair on specific sensitive features such as gender or age. In this paper, we dive into the extent of unfairness exhibited by LLM-based recommender models based on both T5 and LLaMA backbones, and discuss appropriate methods for promoting equitable treatment of users in LLM-based recommendation models. We introduce a novel Counterfactually-Fair-Prompt (CFP) method towards Unbiased Foundation mOdels (UFO) for fairness-aware LLM-based recommendation. 
Experiments are conducted on two real-world datasets, MovieLens-1M and Insurance, and compared with both matching-based and sequential-based fairness-aware recommendation models. Results show that CFP achieves better recommendation performance with a high level of fairness. Source code is anonymously released for reproducibility\footnote{Code and data: \small{https://github.com/agiresearch/UP5}}. 
\end{abstract} 




\maketitle

\section{Introduction} 
\begin{figure}[t]
    \centering
    \includegraphics[scale=0.45]{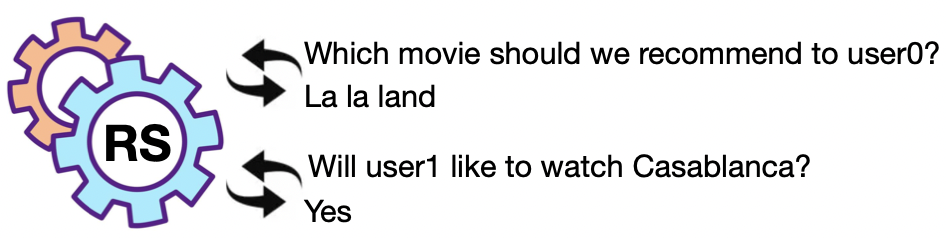}
    \caption{Toy examples of the input-output for prompt-driven LLM-based recommendation models.}
    \label{fig:llmrs_example}
    \vspace{-10pt}
\end{figure}

Large Language Model (LLM) has revolutionized the research in NLP \cite{gpt-3, gpt4}, and its application on Recommender Systems (RS) also attracts soaring interest \cite{fan2023recommender,li2023large,chen2023large,lin2023can,liu2023pre}.
Recommender Systems \cite{RSsurvey} are algorithms designed to personalize contents or items for individual users based on their preferences. Through personalized natural language prompts \cite{P5}, Large Language Models can serve as a backbone for RS (LLM4RS) to generate personalized recommendations based on user and item information. Figure \ref{fig:llmrs_example} shows a toy input-output example of prompting LLM-based recommender systems for personalized recommendation.

This paper delves into the fairness of LLM-based recommendation, a significant concern of RS due to its influence on individual decision-making \cite{li2023fairness, unifying, longtermfairness, flexiblefairness, connection, fairnessrecretrieval, algorithmicfairness}. Specifically, we aim to address user-side counterfactual fairness \cite{userfairness, userfairnesstransparency, userexperiments, personalized, twoside} in RS. We ensure that the RS generates recommendations without factoring in the sensitive attributes that users wish to remain undisclosed. For instance, in a movie recommender system, users may seek recommendations that are not influenced by sensitive attributes such as race, gender, or age. For example, an elderly user may also want to watch younger generation movies to catch up with the times, and thus the user does not want to be discriminated on their age in terms of movie recommendation. 
As a result, recommender systems should allow users to convey their sensitive preferences and consider these criteria for generating recommendations, rather than solely relying on the recommendation model's determination.

In traditional RS, each user is modeled either as a single embedding (in matching models) \cite{PMF, zhang2017joint, BiasedMF, DMF, deepmodel, koren2009matrix} such that whether an item should be recommended is computed by the similarity between item embedding and user embedding, or as a sequence of item embeddings from the user's interaction history (in sequential models) \cite{rnntopk, sasrec, bert4rec, session, rrn, dynamic} such that the model will generate the next item based on the history. 
However, in the context of LLM-based recommendation, the user's information is not consolidated into a singular user embedding or a sequence of item embeddings, thus rendering traditional methods inapplicable. As a result, this paper explores methods to remove sensitive information from LLM-based recommendation models for fairness-aware recommendation. Since LLM-based recommendation models contain a large number of parameters storing a rich amount of knowledge for both language understanding and personalized recommendation, to remove unfairness from such models, three challenges need to be addressed: 1) efficient training and inference of the attribute-specific fairness-aware models for each sensitive attribute and their combinations, 2) avoiding training separate models for each combination of sensitive attributes due to a potentially exponential growth in attribute combinations, and 3) minimizing performance decrease on recommendations, as user attributes could be important for the recommendation performance.

In this work, we first explore three methods to probe the unfairness of LLM-based recommendation. Then, we present the Counterfactually-Fair-Prompt (CFP) method to mitigate the user-side unfairness and propose a fairness-aware foundation model, wherein sensitive user attributes, such as gender, age, occupation, etc., can be either removed or preserved based on each user's preference. 
We experiment on two datasets which contain sensitive attributes, \textit{MovieLens-1M} and \textit{Insurance}, for fairness research, showing the effectiveness of our model in eliminating unfairness while maintaining a high level of recommendation performance. 


The paper proceeds as follows: Section \ref{section:related} presents an overview of the related work on fairness in LLM and RS;
Section \ref{section:preliminary} briefly introduces the preliminary of LLM-based recommendation and its fairness motivation; 
Section \ref{section:method} introduces the proposed CFP model. Section \ref{section:experiment} presents the experimental results for both single-attribute fairness and combined-attribute fairness. Section \ref{section:ablation} provides ablation studies and hyperparameter sensitivity analysis. Section \ref{section:conclusion} concludes the paper.

\begin{figure*}[ht]
    \centering
    \includegraphics[scale=0.44]{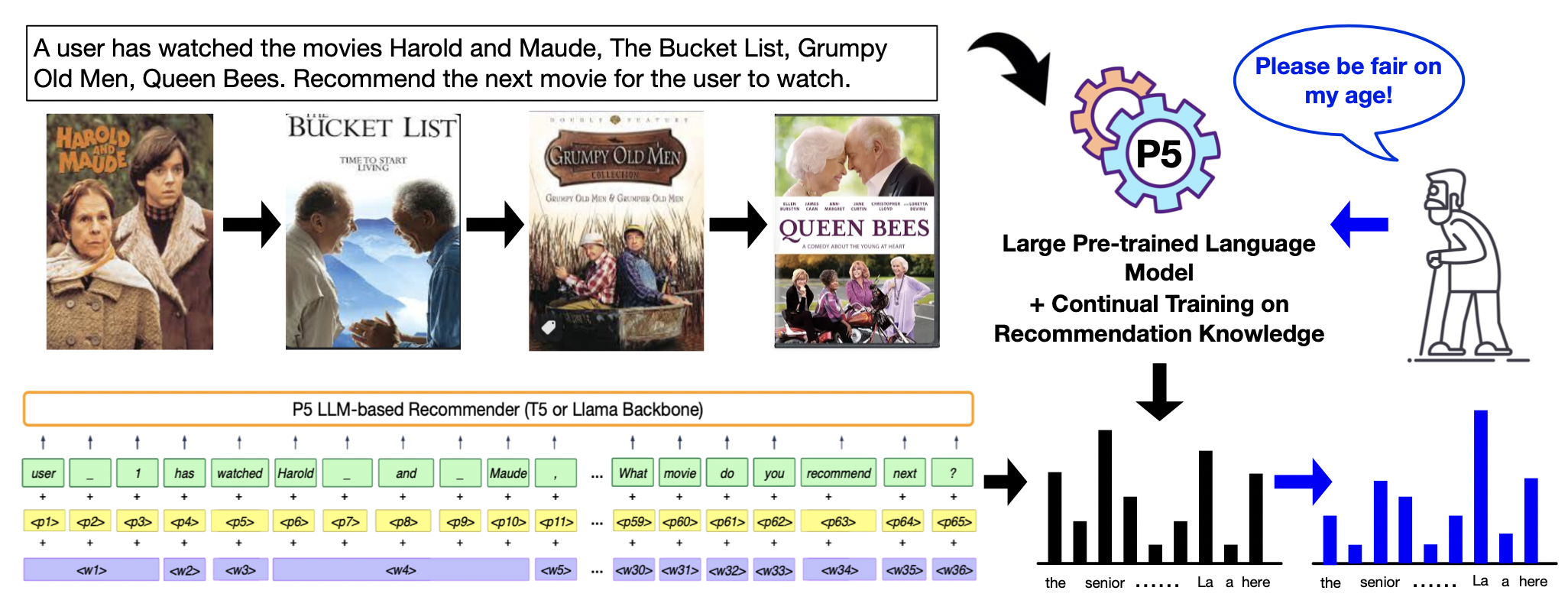}
    \vspace{-10pt}
    \caption{Counterfactual fairness of LLM-based recommendation given the user's choice of sensitive attribute.}
    \vspace{-10pt}
    \label{fig:enter-label}
\end{figure*}

\vspace{-5pt}
\section{Related Work}
\label{section:related}

\paragraph{Fairness of Recommender Systems.}
Since recommender systems involve various stakeholders such as users, item providers, and the platform itself, fairness is a multi-sided concept in recommender systems \cite{li2023fairness,wang2023survey,ekstrand2019fairness}. For user-side fairness, especially counterfactual fairness, it is usually defined as whether recommendations for a user are made independently of the user's sensitive attributes, which is measured by determining whether the recommendation outcomes for a given user are equivalent in both the factual and counterfactual scenarios with respect to a specific attribute \cite{ explainable, counterfactuallearning, personalized}. In the context of RS, a counterfactual world is an alternate scenario in which the user's sensitive attributes are manipulated while all other attributes independent of the sensitive attributes are held constant, as defined in the following \cite{personalized}:


\noindent\textbf{Definition 2.1 (Counterfactually fair recommendation)}
\textit{An RS is counterfactually fair iff. for any possible user $u$ with features $X = x$ and $K = k$, where $K$ are the user's sensitive attributes and $X$ are the attributes that are causally independent of $K$,}
\begin{equation}
    P(L_k|X = x, K=k) = P(L_{k'}|X = x, K = k)
\end{equation}
\textit{holds for all $L$ and any value $k$ attainable by $K$, where $L$ is the recommendation list for user $u$. }

A sufficient condition for RS to be counterfactually fair is to remove the user's sensitive information when generating recommendations
so that the recommendation outcome remains unchanged across various counterfactual scenarios \cite{personalized,selective}, which is ultimately similar to the fairness of language models except that we focus on user representations other than attribute-related words.
\citeauthor{personalized} and \citeauthor{selective} explored personalized counterfactual fairness for traditional RS, where \cite{personalized} is developed for matching-based RS while \cite{selective} is for  sequential-based RS. However, counterfactual fairness for LLM-based RS has largely been unexplored, which has unique challenges to solve as we mentioned before. 
Furthermore, existing methods are not directly applicable to LLM-based recommendation.
For example, \citeauthor{personalized} 
requires updating all parameters in the model for each feature, which is not parameter-efficient and thus unsuitable for large language models.
\citeauthor{selective} appends a prefix prompt and an adapter to the model for improving fairness on sequential recommendation. However, for each attribute combination, a new prefix prompt and a new adapter must be trained from scratch, and thus the method cannot properly handle the exponential combination of attributes. As a result, developing fairness-aware methods for LLM-based recommendation is highly needed.


\paragraph{Fairness of Large Language Models.}
Fairness of language models is usually concerned with whether embeddings for attribute-related words such as gender-related words are associated with stereotypes \cite{nullitout}. Recent studies have highlighted the potential of unfairness in the pre-training data of LLMs, which leads to the generation of harmful or offensive content, including discrimination against marginalized groups. Consequently, there has been an increased research focus on addressing the harmfulness issues of LLMs, with a particular emphasis on unfairness. In a study conducted by \citeauthor{zhuo2023exploring}, the fairness of LLMs was examined using two datasets specifically designed to assess bias in the context of general question answering and text generation tasks. Another research effort by \citeauthor{sun2023safety} evaluated the safety of Chinese LLMs, including an examination of fairness. The study involved observing the frequency of harmful information present in the responses generated by LLMs. This approach provided insights into the potential unfairness and its impact on the safety of these models.
\cite{zhang2023chatgpt} and \cite{li2023fairnesschatgpt} tested the fairness of ChatGPT on recommendation, education, medical and legal tasks, though they did not provide solutions for the unfairness problems.
There also exist several benchmark datasets that are used to better evaluate the unfairness and other harmfulness of LLMs, such as RedTeamingData \cite{ganguli2022red} and HELM \cite{liang2022holistic}.
While there have been numerous investigations into the fairness of LLMs within the field of NLP, there is currently a gap of research in terms of addressing the fairness problems of LLM-based recommender systems.

\section{Preliminary of LLM-based Recommendation}
\label{section:preliminary}
Foundation Models such as Large Language Lodels (LLMs), e.g., BERT \cite{bert}, Llama \cite{llama}, T5 \cite{t5}, and GPT-3 \cite{gpt-3}, have been shown to effectively learn rich semantics from web-scale data and transfer knowledge in pre-training data to various downstream NLP tasks. 
For recommender systems, P5 \cite{P5,xu2023openp5} stands as a seminal framework for foundational recommendation models, grounded in the architecture of LLM backbone models, including both encoder-decoder configuration T5 \cite{t5} and decoder-only model Llama \cite{llama}. By integrating various recommendation tasks---ranging from item generation, recommendation explanation, to rating prediction---P5 enhances the adaptability of contemporary recommendation methodologies. 

In our research, we employ both T5 \cite{t5} and OpenLlama \cite{openlm2023openllama} backbones within the P5 framework to execute experiments targeting unfairness mitigation. In this particular section, we train P5 and probe its fairness problem to motivate the fairness research for LLM-based recommendation. More specifically, we train P5 on two tasks: direct recommendation and sequential recommendation.
Direct recommendation generates recommendations without any user-item interaction history in the input prompt, while sequential recommendation explicitly involves user-item interaction histories. We use the simple and effective sequential ID indexing method for both tasks \cite{hua2023index}.
The prompt for each task is presented in the following square box.
\vspace{-10pt}
\begin{formal}
\small
\textbf{\textit{Direct Recommendation}}\\
\textbf{Input}: Which movie user\_\{\{user\_ID\}\} would like to watch among the following candidates? \{\{List of 100 candidate movies\}\}. \textbf{Output}: \{\{movie\_ID\}\}\\
\textbf{\textit{Sequential Recommendation}}\\
\textbf{Input}: User\_\{\{user\_ID\}\} has already watched the following movies \{\{the sequence of movie IDs this user watched\}\}. Which movie user\_\{\{user\_ID\}\} would like to watch next? \textbf{Output}: \{\{movie\_ID\}\}
\end{formal}
\vspace{-10pt}

\paragraph{Motivating Fairness Concerns.} As presented above, P5 does not explicitly involve any sensitive-attribute-related textual description for users. However, it can still implicitly infer user sensitive attributes and possibly use it for recommendation, even though users may not want to include such sensitive attributes when generating recommendations for them. We use three methods for probing the user attributes from the LLM: 1) eliciting the attributes through in-context learning based on manually designed prompts, 2) generating attributes by tuning soft probing prompts, and 3) training a classifier on the embeddings corresponding to the user tokens in the input.

\begin{figure}[t]
    \centering
    \includegraphics[scale=0.2]{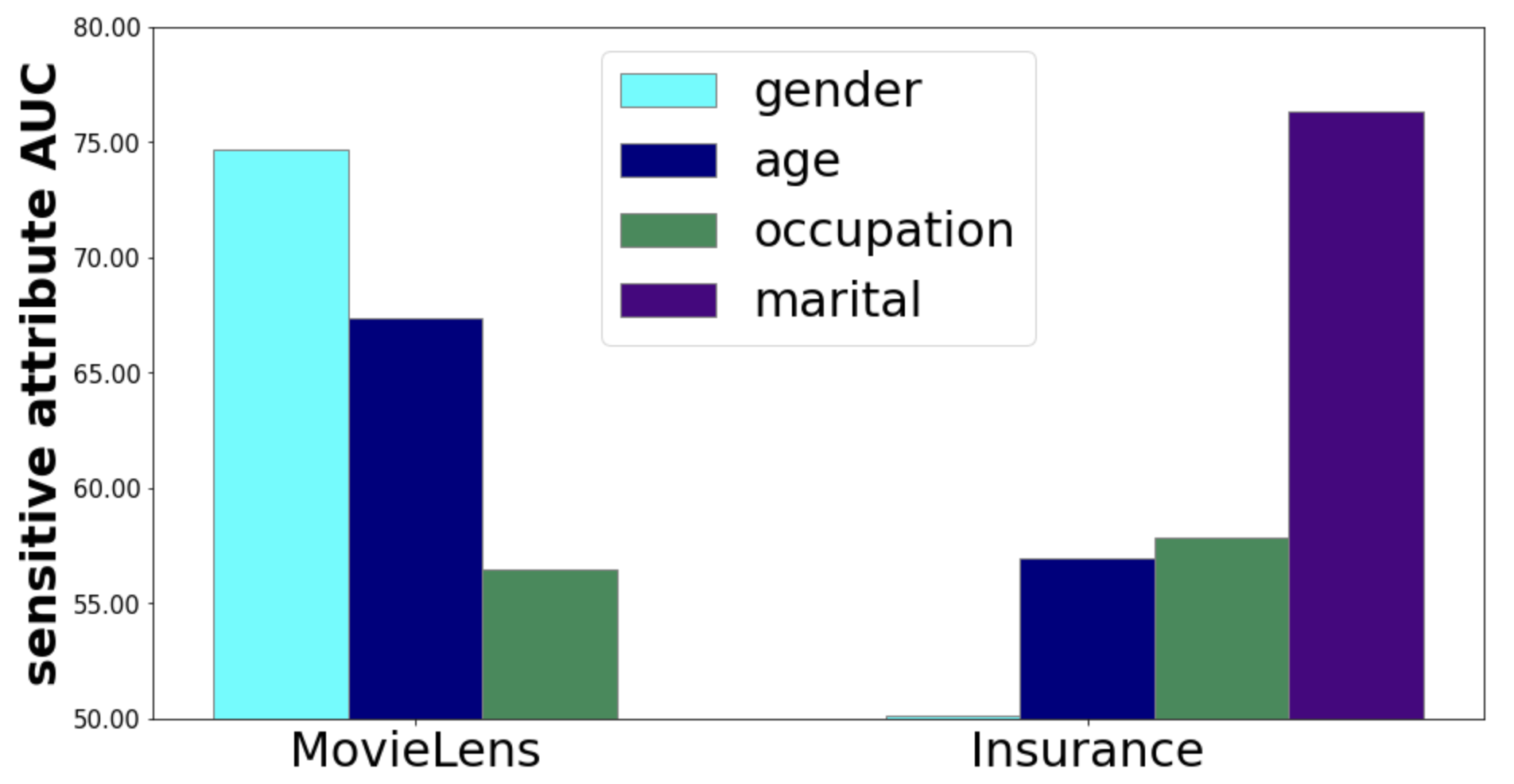}
    \vspace{-10pt}
    \caption{Inferring sensitive attribute information from LLM-based recommendation model.}
    \label{fig:auc}
\vspace{-15pt}
\end{figure}

Figure \ref{fig:auc} presents the AUC score of predicting the sensitive attributes (gender, age, occupation, and marital\_status) from the LLM on Movielens and Insurance datasets, while more details of the implementation and results are presented in the Appendix \ref{appendix:probing}. Experimental results show that both the soft prompt tuning and the classification methods can detect user-sensitive attributes from the LLM, though manual prompts fail. The classification and soft prompt tuning methods both generate above-random predictions on user attributes. This result implies that even though the training and tuning process of LLM-based recommendation does not directly involve users' sensitive attributes, such sensitive information is still inferred by the LLM and embeded in the LLM parameters for generating recommendations, though users may not want their recommendations to be influenced by certain sensitive attributes. As a result, it is important to develop sensitive mitigation methods so as to enable counterfactually fair LLM-based recommendation, which we will introduce in the following sections.

\section{Counterfactually-Fair Prompting}
\label{section:method}

We propose a Counterfactually-Fair-Prompt (CFP) method to mitigate the unfairness of LLM-based recommendation, resulting in the development of a fair and accurate recommendation foundation model. Our approach is 1) personalized, since each user can choose the attributes that they wish to be treated fairly on, and 2) space and time efficient, since our approach does not require retraining the entire foundation model and only needs to train the prefix prompts. The key idea of the CFP method is to train a counterfactually-fair prompt (CFP): For encoder-decoder LLM, we need an encoder prompt $p_{enc}$ to remove sensitive attributes and a decoder prompt $p_{dec}$ to preserve the model performance; For decoder-only LLM, we only need a decoder prompt. Our goal is to learn such CFP so that sensitive information in the user token embeddings is removed by simply concatenating the CFP with the original input prompt.

\begin{figure*}
    \centering
    \includegraphics[width=0.98\textwidth]{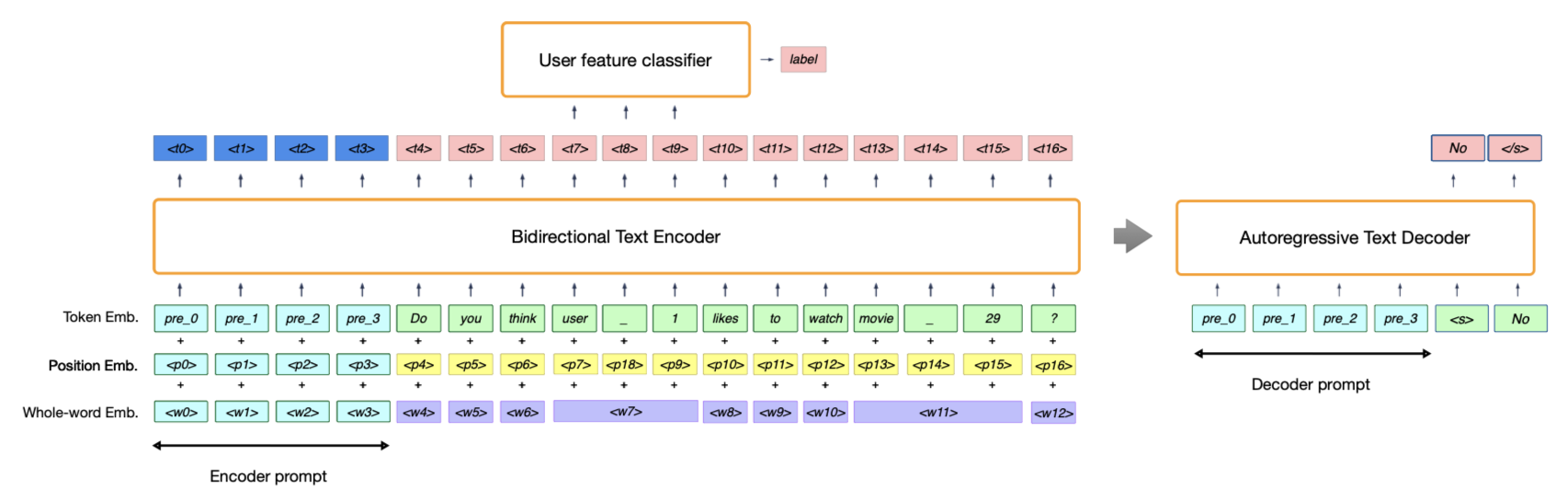}
    \vspace{-15pt}
    \caption{Counterfactually-Fair-Prompting method for sensitive attribute mitigation and fairness improvement}
    \label{fig:UP5}
    \vspace{-10pt}
\end{figure*}

CFPs are trained by adversarial learning \cite{adversarial2005, adversarial2018, adversarial2022}.
Adversarial learning requires a discriminator module \cite{deeplearningadversarial} aiming at precise extraction of attribute values from embeddings, while CFP aims at obfuscation of the discriminator's efforts. Thus, the stronger the discriminator, the more effectively we can clean sensitive information from embeddings.
According to the probing experiments in Section \ref{section:preliminary}, the multi-class classifier is a stronger prober than other approaches. Thus, we utilize the classifier as the discriminator in adversarial learning. Figure \ref{fig:UP5} shows the model architecture. We also present the results of using the soft probing prompt as a discriminator in Section \ref{soft_probing_prompt_as_discriminator} for comprehensiveness.

The model training involves an iterative process where the CFP and the classifier are optimized in succession. For each attribute $k$, we denote the recommendation loss as $L_{rec}^k$ and the discriminator loss as $L_{dis}^k$. Let $\mathcal{M}$ denote the recommendation foundation model and $\mathcal{C}_k$ as the classifier.
$L_{rec}^k$ is a negative log-likelihood loss that encourages generating the correct item $y$:
\begin{equation}
\small
\label{rec_loss}
    L_{rec}^k = -\sum\nolimits_{j=1}^{|y|}\log P(y_j|p_{enc_{k}}\circ x, p_{dec_{k}}\circ y_{0:j-1}, \mathcal{M})
\end{equation}
$L_{dis}^k$ is a Cross-Entropy Loss (CEL) that encourages predicting user attribute $k$ correctly based on the average of user-relevant token embeddings $\mathcal{E}$ (e.g., the tokens ``user'', ``\_'', and ``1'' in Figure \ref{fig:UP5}) conditioned on $p_{enc_k}$. Denoting $u$ as the user, and $c_u$ the correct attribute value for the user, $L_{dis}$ is:
\begin{equation}
\label{dis_loss}
L_{dis}^k = \text{CEL}(c_u, \mathcal{C}_k(\text{mean}(\mathcal{E}_u)))
\end{equation}
The adversarial loss $L_k$ for each attribute $k$ is defined as below, where $\lambda_k$ denotes the discriminator weight for attribute $k$:
\begin{equation}
\label{ad_loss}
    L_k = \sum\nolimits_u L_{rec}^k - \lambda_k \cdot L_{dis}^k
\end{equation}
The training algorithm is presented in Appendix \ref{algorithm}.



\subsection{Prompt Mixture}
Users may seek recommendations that remain impartial to several attributes at the same time. For instance, they may want a model to overlook details like gender and marital status but still value recommendations that resonate with movie preferences typical for their age group. Consequently, CFPs must possess the capacity to exclude several attributes in tandem. An elementary approach might involve developing a prompt for every possible attribute combination, but this is operationally taxing given the exponential growth in the number of combinations.

\begin{figure}[ht]
    \centering
    \includegraphics[scale=0.45]{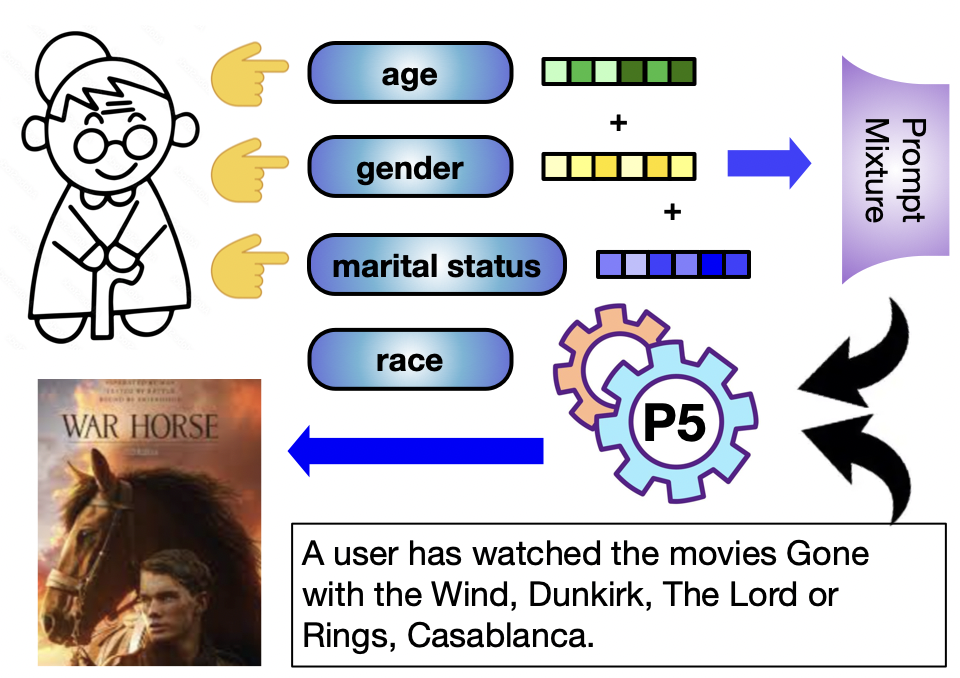}
    \caption{Prompt Mixture over CFPs from 3 attributes}
    \label{fig:prompt_mixture}
    \vspace{-10pt}
\end{figure}

To solve the challenge, we propose a Prompt Mixture (PM) module. This module comprises a singular attention layer that combines the embeddings from various single-attribute CFPs to integrate user preferences. The attentional framework offers flexibility regarding input length, allowing for the integration of a variable number of CFPs, each potentially of distinct lengths. The PM is adept at processing information from different CFPs, masking sensitive user information while preserving other relevant details within the model-generated hidden states. This positions the PM as an invaluable instrument for a user-controllable LLM-based recommendation model since users have the freedom to choose different sensitive feature combinations, facilitating the assimilation of multifaceted user stipulations without the necessity for specialized model training for each unique combination of requirements (Figure \ref{fig:prompt_mixture}). 

Similar to single attribute prompt learning introduced above, PM is also trained based on adversarial learning, where each optimization step includes a random combination of sensitive attributes selected to be removed. PM  takes a concatenation of multiple single-attribute prefix prompts as input and generates a new prompt, which is optimized to simultaneously decrease the recommendation loss and increase the sum of discriminator loss of multiple classifiers. 
The loss function for one step with a set of randomly selected attributes $\mathbf{K}$ is:
\begin{equation}
\label{eq:multi-attribute-loss}
L_{\mathbf{K}} = \sum\nolimits_u (L_{rec}^{\mathbf{K}} - \Sigma_{k\in\mathbf{K}}\lambda_k \cdot L_{dis}^k)
\\
\\
\\
\end{equation}



\section{Experiments}
\label{section:experiment}

This section presents the experimental results of CFP on a variety of metrics, including recommendation performance and fairness level. The results show the model's ability to achieve fairness in both single-attribute and multi-attribute scenarios.


\subsection{Experimental Setup}
\paragraph{\bf Datasets} Experiments are conducted on the MovieLens-1M dataset and Insurance dataset:\\
\textbf{MovieLens-1M}\cite{ml}: The dataset contains user-movie interactions and user profile information: gender, age, and occupation. Gender is a binary feature, occupation is a twenty-one-class feature,
and age is a seven-class feature.
\textbf{Insurance}\footnote{https://www.kaggle.com/datasets/mrmorj/insurance-recommendation}: The dataset contains user-insurance interactions. The user profile contains four features: gender, marital status, age, and occupation. Gender is a binary feature, marital status is a seven-class feature, occupation is a six-class feature, and age is a five class feature.

\paragraph{\bf Evaluation Metrics}
To evaluate direct recommendation and sequential recommendation tasks, one correct item is predicted among 100 randomly selected negative samples for both tasks. The metrics are Hit@$k$ for $k$ in $\{1,3,10\}$. We adopt the commonly used leave-one-out strategy (for each user, treat the second-to-last interacted item to be the validation item and the last interacted item to be the test item) to create the training, validation, and test datasets. We adopt AUC for user attribute classification to evaluate whether sensitive attributes are involved in recommendations.

\begin{table}[t]
    \centering
    \small
    \setlength{\tabcolsep}{1.5pt}
    \begin{tabular}{l|c|c|c|c|c|c}
    \toprule
    \textbf{Dataset} & \multicolumn{3}{c|}{\textbf{MovieLens}} & \multicolumn{3}{c}{\textbf{Insurance}}\\
        \hline 
        \textbf{Model} & PMF & SimpleX & P5 & PMF & SimpleX & P5 \\
         \hline
        $\uparrow$ Hit@1 & 19.91 & 17.94 & \textbf{20.57} & 70.20 & 76.50 & \textbf{82.53} \\
        $\uparrow$ Hit@3 & 38.66 & \textbf{38.79} & 38.38 & 75.23 & 80.12 & \textbf{92.68} \\
        $\uparrow$ Hit@10 & 65.69 & 65.69 & \textbf{67.31} & 90.04 & 91.41 & \textbf{98.89}\\
        \hline
        $\downarrow$ AUC (G) & 80.22 & 75.52 & \textbf{74.71} & 52.04 & 53.34 & \textbf{50.11}\\
        $\downarrow$ AUC (A) & 82.37 & 79.39 & \textbf{67.40} & 57.94 & 56.87 & \textbf{50.09}\\
        $\downarrow$ AUC (O) & 61.32 & 59.40 & \textbf{56.50} & 58.25 & 57.12 & \textbf{53.28} \\
        $\downarrow$ AUC (M) & -- & -- & -- & 71.30 & \textbf{68.85} & 69.25 \\
        \bottomrule
    \end{tabular}
    \vspace{-10pt}
    \caption{Results of matching-based recommendation, G means Gender, A means Age, O means Occupation, and M means Marital Status (\%).}
    \label{tab:matching}
    \vspace{-10pt}
\end{table}

\begin{table}[t]
\small
    \centering
    \setlength{\tabcolsep}{3pt}
    \begin{tabular}{l|c|c|c|c|c|c}
    \toprule
    \textbf{Dataset} & \multicolumn{3}{c|}{\textbf{MovieLens}} & \multicolumn{3}{c}{\textbf{Insurance}}\\
    \hline
        \textbf{Model} & SAS & BERT & P5 & SAS & BERT & P5 \\
         \hline
        $\uparrow$ Hit@1 & 28.39 & 29.30 & \textbf{30.34} & 77.26 & 81.20 & \textbf{84.56}\\
        $\uparrow$ Hit@3 & \textbf{53.89} & 49.06 & 49.26 & 85.15 & 93.33 & \textbf{93.99}\\
        $\uparrow$ Hit@10 & \textbf{76.32} & 70.06 & 67.40 & 95.76 & 98.78 & \textbf{98.98}\\
        \hline
        $\downarrow$ AUC (G) & 91.90 & 78.52 & \textbf{74.71} & 73.23 & 61.20 & \textbf{50.13}\\
        $\downarrow$ AUC (A) & 92.06 & 73.35 & \textbf{67.40} & 57.93 & \textbf{54.34} & 56.92 \\
        $\downarrow$ AUC (O) & 76.57 & 64.79 & \textbf{56.50} & 88.04 & \textbf{54.30} & 57.87\\
        $\downarrow$ AUC (M) & -- & -- & -- & 76.61 & \textbf{76.11} & 76.37\\
        \bottomrule
    \end{tabular}
    \vspace{-10pt}
    \caption{Results of sequential recommendation, G is Gender, A is Age, O is Occupation, and M is Marital Status (\%). SAS is SASRec and BERT is Bert4Rec.}
    \label{tab:sequential}
    \vspace{-15pt}
\end{table}

\paragraph{\bf LLM Backbone}
We train the LLM recommendation model under the P5 paradigm \cite{P5} using both T5-Base \cite{t5} and OpenLlama-3B \cite{openlm2023openllama} backbones. We present results based on T5 in this section as the main results for comparison, and detailed results for the OpenLlama experiments are presented in the Appendix \ref{llama_result}.

\begin{table*}
\small
    \centering
    \setlength{\tabcolsep}{1pt}
    \resizebox{\linewidth}{!}{
    \begin{tabular}{l|c|c|c|c|c|c|c|c|c|c|c|c|c|c|c|c|c|c}
    \toprule
        \textbf{Dataset} & \multicolumn{9}{c|}{\textbf{MovieLens}} & \multicolumn{9}{c}{\textbf{Insurance}}\\
        \hline 
        \textbf{Attribute} & \multicolumn{3}{c|}{\textbf{Gender}} & \multicolumn{3}{c|}{\textbf{Age}} & \multicolumn{3}{c|}{\textbf{Occupation}} & \multicolumn{3}{c|}{\textbf{Age}} & \multicolumn{3}{c|}{\textbf{Marital}} & \multicolumn{3}{c}{\textbf{Occupation}} \\
        \hline
        \textbf{Model} & C-PMF & C-SX & CFP & C-PMF & C-SX & CFP & C-PMF & C-SX & CFP & C-PMF & C-SX & CFP & C-PMF & C-SX & CFP & C-PMF & C-SX & CFP\\
         \hline
        $\uparrow$ Hit@1 & \textbf{16.73} & 13.96 & 16.38 & 17.42 & 13.87 & \textbf{21.22} & 15.60 & 14.06 & \textbf{21.00} & 67.61 & 71.14 & \textbf{82.53} & 66.68 & 71.50 & \textbf{81.03} & 68.51 & 71.09 & \textbf{82.53} \\
        $\uparrow$ Hit@3 & 34.03 & 29.56 & \textbf{35.04} & 34.20 & 29.61 & \textbf{39.22} & 34.36 & 29.56 & \textbf{38.50} & 73.25 & 83.23 & \textbf{92.68} & 74.23 & 83.00 & \textbf{90.58} & 74.09 & 82.23 & \textbf{92.68}\\
        $\uparrow$ Hit@10 & 65.32 & 56.02 & \textbf{65.82} & 65.18 & 55.42 & \textbf{67.30} & 65.33 & 56.02 & \textbf{69.49} & 85.98 & 92.65 & \textbf{98.89} & 85.99 & 96.50 & \textbf{97.66} & 85.95 & 93.27 & \textbf{98.89}\\
        $\downarrow$ AUC & 56.62 & 70.80 & \textbf{54.19} & 62.55 & 79.26 & \textbf{52.91} & 56.01 & 57.02 & \textbf{50.00} & 50.81 & 51.26 & \textbf{50.09} & \textbf{52.10} & 56.23 & 52.19 & 54.40 & 
        \textbf{52.09} & 53.28\\
        \bottomrule
    \end{tabular}
    }
    \vspace{-10pt}
    \caption{Results of single-attribute fairness-aware prompting on matching-based models (\%)}
    \label{tab:debias_matching}
    \vspace{-10pt}
\end{table*}

\begin{table*}
\small
    \centering
    \setlength{\tabcolsep}{1pt}
    \resizebox{\linewidth}{!}{
    \begin{tabular}{l|c|c|c|c|c|c|c|c|c|c|c|c|c|c|c|c|c|c}
    \toprule
        \textbf{Dataset} & \multicolumn{9}{c|}{\textbf{MovieLens}} & \multicolumn{9}{c}{\textbf{Insurance}}\\
        \hline 
        \textbf{Attribute} & \multicolumn{3}{c|}{\textbf{Gender}} & \multicolumn{3}{c|}{\textbf{Age}} & \multicolumn{3}{c|}{\textbf{Occupation}} & \multicolumn{3}{c|}{\textbf{Age}} & \multicolumn{3}{c|}{\textbf{Marital}} & \multicolumn{3}{c}{\textbf{Occupation}} \\
        \hline
        \textbf{Model} & S-SAS & S-B4 & CFP & S-SAS & S-B4 & CFP & S-SAS & S-B4 & CFP & S-SAS & S-B4 & CFP & S-SAS & S-B4 & CFP & S-SAS & S-B4 & CFP\\
         \hline
        $\uparrow$ Hit@1 & 20.87 & 23.48 & \textbf{26.82} & 22.95 & 27.98 & \textbf{31.23} & 18.90 & 24.33 & \textbf{31.66} & 69.40 & 81.20 & \textbf{82.08} & 70.10 & 75.33 & \textbf{80.63} & 70.09 & 81.20 & \textbf{82.62} \\
        $\uparrow$ Hit@3 & 41.64 & 42.09 & \textbf{45.18} & 44.10 & 49.32 & \textbf{51.18} & 20.84 & 43.29 & \textbf{50.73} & 80.05 & \textbf{93.33} & 92.62 & 80.38 & 84.54 & \textbf{90.16} & 80.38 & \textbf{93.33} & 92.65\\
        $\uparrow$ Hit@10 & 60.82 & 62.43 & \textbf{64.38} & 66.00 & 69.38 & \textbf{67.70} & 43.87 & 59.74 & \textbf{67.45} & 88.34 & \textbf{98.78} & 98.37 & 88.49 & 94.34 & \textbf{98.38} & 88.91 & \textbf{98.78} & 98.54\\
        $\downarrow$ AUC & 59.72 & 58.33 & \textbf{54.19} & 60.20 & 67.33 & \textbf{52.91} & 67.27 & 60.36 & \textbf{50.00} & 57.48 & 53.34 & \textbf{51.23} & 66.51 & 69.11 & \textbf{50.03} & 86.66 & 54.30 & \textbf{50.82}\\
        \bottomrule
    \end{tabular}
    }
    \vspace{-10pt}
    \caption{Results of single-attribute fairness-aware prompting on sequential models (\%)}
    \label{tab:debias_sequential}
        \vspace{-10pt}
\end{table*}

\paragraph{Baselines}
We adopt four SOTA fairness-aware models as baselines: \citeauthor{personalized}'s Counterfactual-filter method over PMF (C-PMF) and SimpleX (C-SX), and \citeauthor{selective}'s Selective-prompt-adapter method on SASRec (S-SAS) and BERT4Rec (S-B4).
PMF \cite{mnih2007probabilistic,PMF} is the Probabilistic Matrix Factorization model that adds Gaussian prior into the user and item latent factor distributions for matrix factorization. 
SimpleX \cite{simplex} is a contrastive learning model based on cosine contrastive loss which has achieved state-of-the-art performance on recommendation performance. 
\citeauthor{personalized}'s unfairness-removing filters are applied right after the user embedding computed by PMF and SimpleX, which creates C-PMF and C-SX.
SASRec
\cite{sasrec} is a sequential recommendation model based on left-to-right self-attention mechanism.
BERT4Rec
\cite{bert4rec} is a bidirectional sequential recommendation model based on BERT. 
\citeauthor{selective}'s prompts are appended to item sequences and adaptors are inserted into each Transformer encoder block in SASRec and BERT4Rec, which creates S-SAS and S-BERT.

\begin{table*}
\small
    \centering
    \setlength{\tabcolsep}{5pt}
    \begin{tabular}{l|c|c|c|c|c|c|c|c|c|c|c|c}
    \toprule
    \textbf{Model} & \multicolumn{3}{c|}{\textbf{GA}} & \multicolumn{3}{c|}{\textbf{GO}} & \multicolumn{3}{c|}{\textbf{AO}} & \multicolumn{3}{c}{\textbf{GAO}} \\
    \hline 
        \textbf{Attribute} & C-PMF & C-SX & CFP & C-PMF & C-SX & CFP & C-PMF & C-SX & CFP & C-PMF & C-SX & CFP \\
        \hline
        $\uparrow$ Hit@1 & 14.93 & 15.61 & \textbf{16.33} & 15.25 & 15.53 & \textbf{18.67} & 14.84 & 15.43 & \textbf{21.37} & 15.09 & 15.67 & \textbf{20.18} \\
        $\uparrow$ Hit@3 & 32.11 & 31.79 & \textbf{37.48} & 32.70 & 31.84 & \textbf{39.02} & 31.83 & 31.87 & \textbf{39.83} & 32.58 & 31.85 & \textbf{38.79} \\
        $\uparrow$ Hit@10 & 60.51 & 58.82 & \textbf{66.89} & 60.58 & 58.78 & \textbf{66.39} & 59.51 & 58.71 & \textbf{68.40} & 60.75 & 58.87 & \textbf{66.78} \\
        $\downarrow$ Avg. AUC & 58.03 & 70.25 & \textbf{54.22} & 56.57 & 60.90 & \textbf{52.10} & 56.57 & 64.41 & \textbf{50.00} & 56.54 & 65.19 & \textbf{53.21} \\
        \bottomrule
    \end{tabular}
    \vspace{-10pt}
    \caption{Results of multi-attribute fairness-aware prompting on MovieLens dataset (\%)}
    \label{tab:combine_debias_matching_ml}
    \vspace{-10pt}
\end{table*}

\begin{table*}
\small
    \centering
    \setlength{\tabcolsep}{5pt}
    \begin{tabular}{l|c|c|c|c|c|c|c|c|c|c|c|c}
    \toprule
    \textbf{Model} & \multicolumn{3}{c|}{\textbf{AO}} & \multicolumn{3}{c|}{\textbf{AM}} & \multicolumn{3}{c|}{\textbf{MO}} & \multicolumn{3}{c}{\textbf{AMO}} \\
    \hline 
        \textbf{Attribute} & C-PMF & C-SX & CFP & C-PMF & C-SX & CFP & C-PMF & C-SX & CFP & C-PMF & C-SX & CFP \\
        \hline
        $\uparrow$ Hit@1 & 63.68 & 71.58 & \textbf{79.00} & 62.27 & 71.23 & \textbf{80.91} & 62.44 & 71.11 & \textbf{78.30} & 64.38 & 72.30 & \textbf{81.63} \\
        $\uparrow$ Hit@3 & 70.55 & 80.50 & \textbf{89.22} & 69.78 & 79.18 & \textbf{90.97} & 69.39 & 81.22 & \textbf{88.45} & 70.11 & 81.78 & \textbf{91.52} \\
        $\uparrow$ Hit@10 & 84.88 & 93.61 & \textbf{97.66} & 83.85 & 93.22 & \textbf{98.73} & 84.88 & 93.52 & \textbf{97.33} & 85.90 & 93.35 & \textbf{97.37} \\
        $\downarrow$ Avg. AUC & 58.38 & 55.98 & \textbf{50.80} & 55.60 & 59.97 & \textbf{50.79} & 57.86 & 59.79 & \textbf{50.64} & 57.44 & 58.43 & \textbf{50.74} \\
        \bottomrule
    \end{tabular}
    \vspace{-10pt}
    \caption{Results of multi-attribute fairness-aware prompting on Insurance dataset (\%)}
    \label{tab:combine_debias_matching_insurance}
    \vspace{-10pt}
\end{table*}

\paragraph{Implementation Details}
The model hyper-parameters are selected within the following range: discriminator weight $\lambda\in\{1, 5, 10, 100\}$, prefix length $\in\{5, 15, 30\}$, batch size = 16, number of steps $T\in\{10, 20\}$ to update $\mathcal{C}$ on $L_{dis}$ or prefix prompt $\mathcal{P}$ on $L_{rec}$, number of batches $R\in\{20\}$ to update prefix prompt $\mathcal{P}$ on adversarial loss $L$.

\subsection{Overall Results of the CFP Model}
This subsection presents the overall results.

\paragraph{Overall Performance} 
Table \ref{tab:matching} and Table \ref{tab:sequential} present the recommendation performance and unfairness of the baseline models for direct recommendation and sequential recommendation respectively. The first 3 rows on each table are the recommendation performance and the last 4 rows show the extent of unfairness. From the result, we see that LLM-based recommendation model (P5) performs better than other models on both datasets. 

\paragraph{Single-Attribute Scenario} 
We compare the CFP model with fair matching-based models C-PMF and C-SX in Table \ref{tab:debias_matching} and fair sequential-based models S-SASRec and S-BERT4Rec in Table \ref{tab:debias_sequential}, since both frameworks provide solutions in single-attribute scenarios.
CFP outperforms both fair matching-based and sequential-based models in terms of both AUC and recommendation accuracy. The AUC of CFP is close to 50\%, indicating a high level of fairness since the model is unable to inferr users' sensitive attributes, and the negative impact on recommendation performance is minimal compared to other models. 

\paragraph{Multi-Attribute Scenario}
We also provide experiment results on multi-attribute fairness treatment, as shown in
Table \ref{tab:combine_debias_matching_ml} and Table \ref{tab:combine_debias_matching_insurance}. The attribute row denotes the set of attributes to be removed, where ``G'' represents ``gender,'' ``A'' represents ``age,'' ``O'' represent ``occupation,'' and ``M'' represents ``marital status''. Two or more attributes together such as ``GA'' means that the sensitive attributes need to be removed at the same time. We compare our CFP model with the two matching-based fairness baselines C-PMF and C-SX from \citeauthor{personalized}, since the sequential fairness baselines from \citeauthor{selective} are unable to handle mutiple attributes. We report the recommendation performance and the average AUC for the targeted user attributes in Table \ref{tab:combine_debias_matching_ml} (MovieLens) and Table \ref{tab:combine_debias_matching_insurance} (Insurance).
We can see that our CFP method under prompt mixture is an effective method to combine the single-attribute prefix prompts, achieving fairness and meanwhile maintaining high recommendation performance.

\section{Detailed Analysis}
\label{section:ablation}
This section discusses the effect of different model designs of the CFP method. We experiment on 1) how hyperparameters such as prompt length and discriminator weights affect the performance, and 2) how the choice of discriminator (classifier or soft probing prompt) affects the performance.

\vspace{-5pt}
\paragraph{\bf Hyperparameter Sensitivity}
\begin{figure}[H]
    \centering
    \includegraphics[scale=0.295]{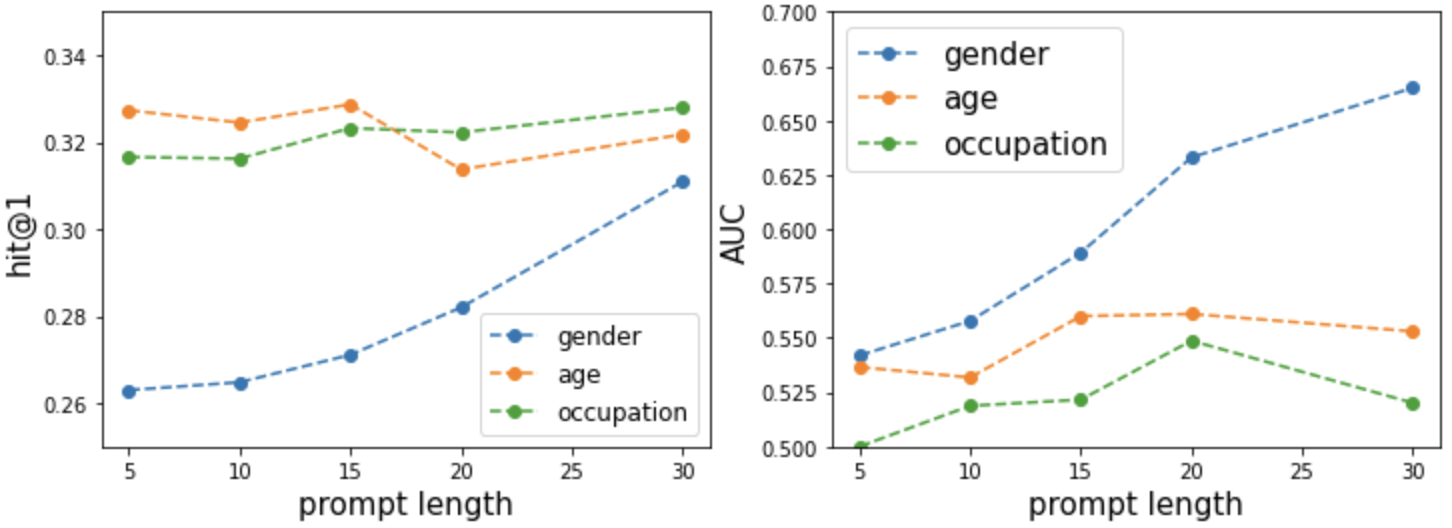}
    \vspace{-10pt}
    \caption{Different prompt length on MovieLens}
    \label{fig:movielens_length}
\vspace{-20pt}
\end{figure}

\begin{figure}[H]
    \centering
    \includegraphics[scale=0.293]{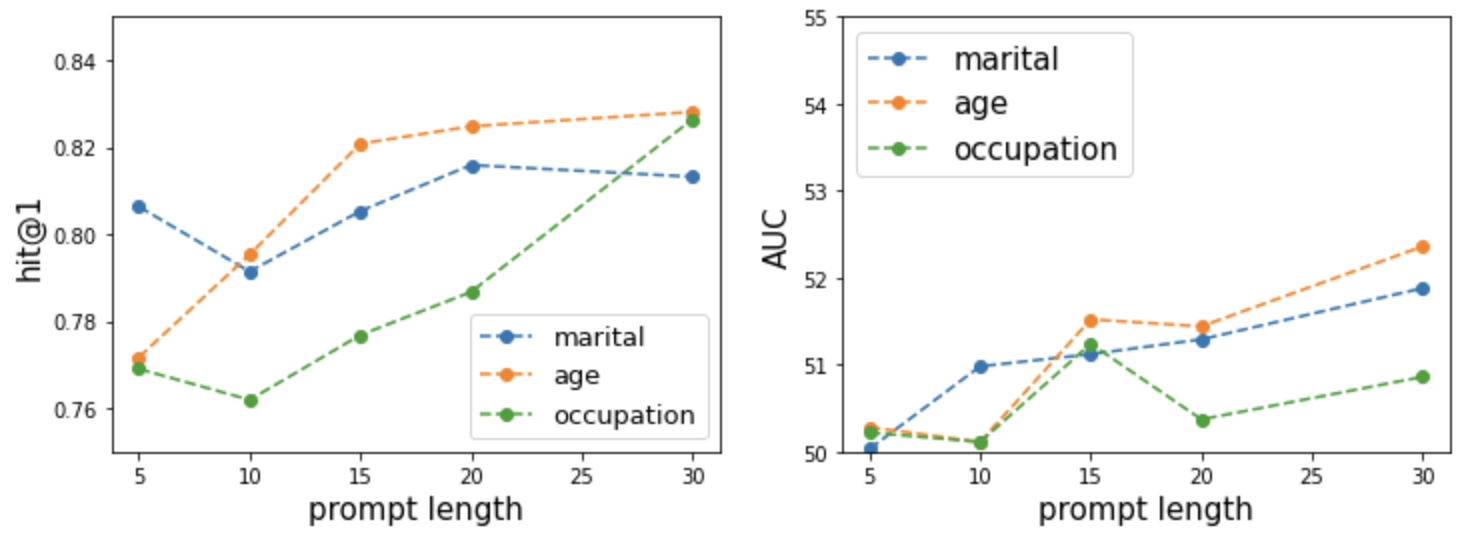}
    \vspace{-10pt}
    \caption{Different prompt length on Insurance}
    \label{fig:insurance_length}
\vspace{-10pt}
\end{figure}

\begin{figure}[H]
    \centering
    \includegraphics[scale=0.3]{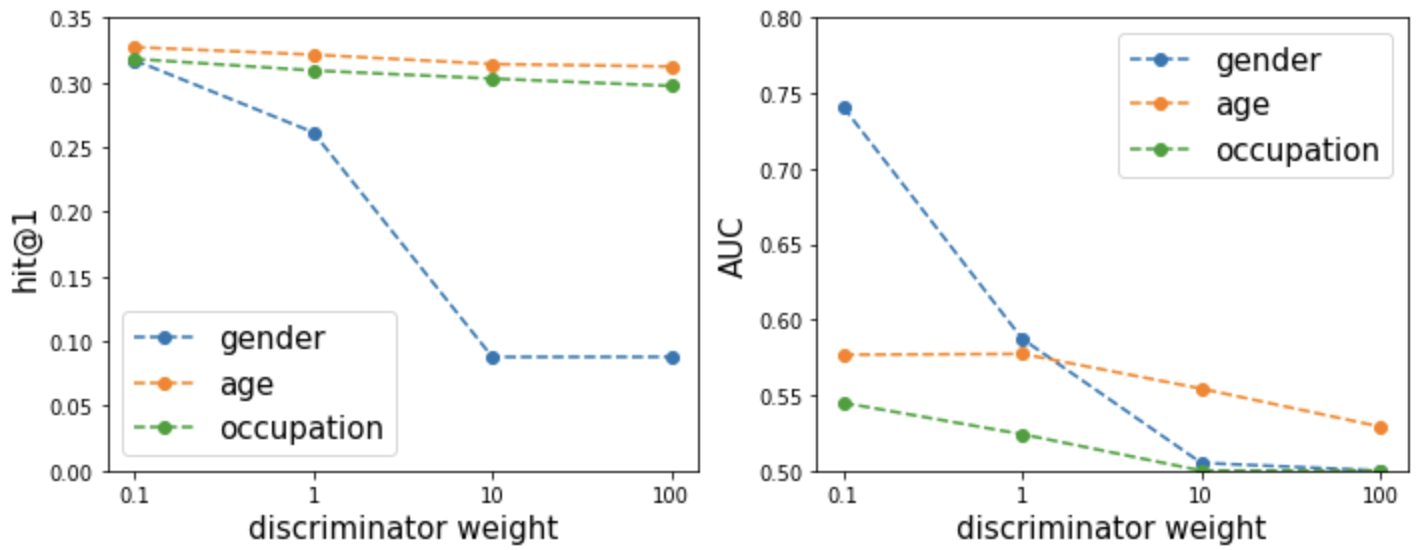}
    \caption{Different discriminator weight on MovieLens}
    \label{fig:movielens_discriminator}
\vspace{-20pt}
\end{figure}

\begin{figure}[H]
    \centering
    \includegraphics[scale=0.30]{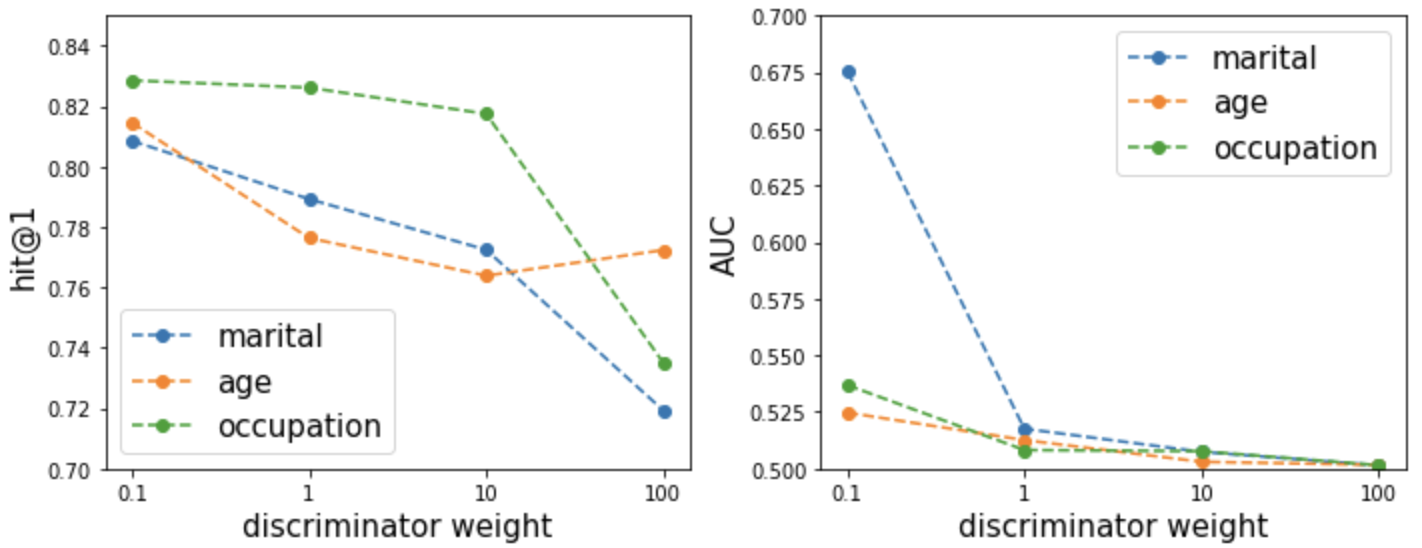}
    \caption{Different discriminator weight on Insurance}
    \label{fig:insurance_discriminator}
\end{figure}

In this section, we study the effect of prompt length (5, 10, 15, 30) and discriminator weight (0.1, 1, 10, and 100) on both recommendation performance (Hit@1 on sequential recommendation) and attribute detection performance (AUC).
Figure \ref{fig:movielens_length} and \ref{fig:insurance_length} present the effects of prefix prompt length on MovieLens and Insurance, respectively. In general, longer prefix length hurts fairness but improves the recommendation performance. 
Figure \ref{fig:movielens_discriminator} and \ref{fig:insurance_discriminator} present the results under different discriminator weight $\lambda$, showing that larger weights bring better fairness but hurt the recommendation performance since the fairness term dominates the loss. Results indicate that we need to choose the prompt length and discriminator weight carefully to balance the fairness-recommendation trade-off. 

\begin{figure}[H]
\centering
\begin{minipage}{.23\textwidth}
  \centering
  \includegraphics[width=1\linewidth]{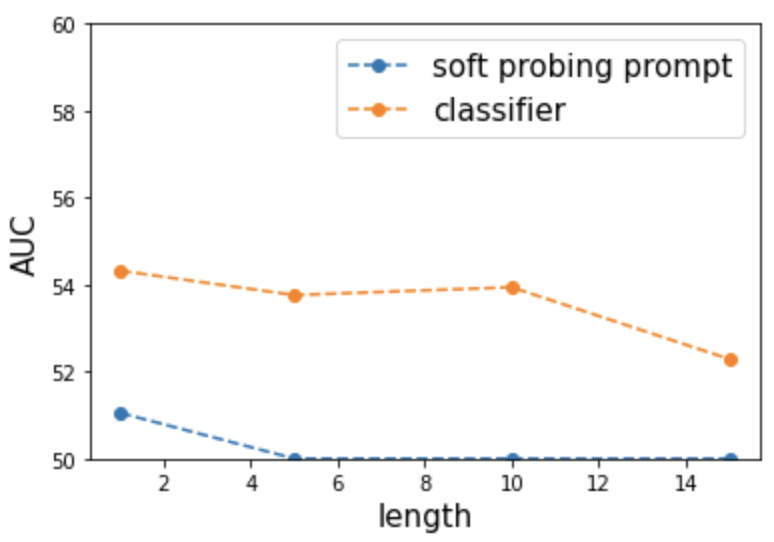}
  \label{fig:test1}
\end{minipage}%
\begin{minipage}{.23\textwidth}
  \centering
  \includegraphics[width=1\linewidth]{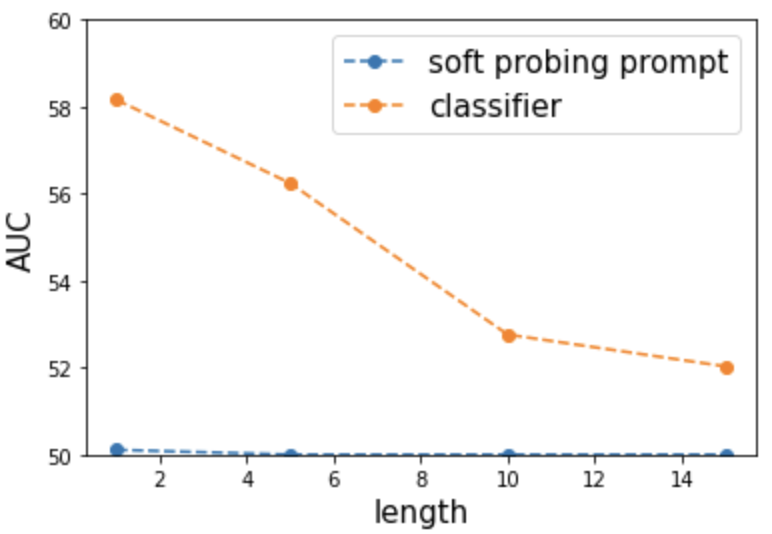}
  \label{fig:test2}
\end{minipage}
\vspace{-20pt}
\caption{Effect of different lengths on AUC using soft probing prompt and classifier for probing}
\label{fig:length}
\end{figure}

\paragraph{Soft Probing Prompt as Discriminator}
\label{soft_probing_prompt_as_discriminator}
This section discusses whether we can use soft probing prompt as the discriminator in adversarial training to improve fairness. According to the motivating experiments on probing fairness of LLMs (Section \ref{section:preliminary}), soft probing prompt is a weaker tool to extract user attribute information compared with multi-class classier. To further validate this, 
we train the CFP using soft probing prompt as the discriminator. To test the effectiveness of the trained prompts, we append the trained CFP in front of the model inputs and then use 1) soft probing prompt and 2) multi-class classifier to extract user attribute information.
We present the results on the Insurance dataset targeting the marital status attribute under different lengths of the CFP in Figure \ref{fig:length}, and other dataset and attributes have similar observations. We see that 1) the probing prompts cannot extract any user attribute since its AUC is close to 50\%, while the classifier can still extract non-trivial sensitive attribute information from the LLM. 2) longer CFPs are more effective in removing sensitive attributes, since the classifier can extract less information, while AUCs for probing prompts are always around 50\%. As a result, this result shows that to train CFPs, it is better to use the classifier instead of soft probing prompt as the discriminator. 


\vspace{-15pt}
\section{Conclusion and Future Work}\label{section:conclusion}
This paper explores the unfairness issue of LLM for recommendation by first probing the unfairness issue of LLM-based recommendation models, and then proposing a novel CFP method to mitigate the issue, enabling a fair recommendation foundation model. In the future, we will explore fairness in other aspects of LLM-based recommendation, such as explanation generation and conversational recommendation.
We are also committed to developing user-friendly interfaces and algorithms that are responsive to user specifications for user controllable fairness without compromising the system's performance or user experience.

\section*{Limitation}
The paper investigates unfairness issues in large language models for recommender systems. However, the paper still has several limitations. In particular, though we explored fairness of LLM-based recommendation over several sensitive features such as gender, age, and occupation, we did not study the bias problems with regard to historically disadvantaged groups. 
The reason is because we are not aware of the availability of any dataset containing such sensitive feature information. In the future, when such dataset becomes available, we plan to extend our exploration on the fairness of LLM-based recommendation over such features.

\section*{Ethical Consideration}
Our method is proposed to increase the fairness of recommendation performance for users. It will unlikely lead to negative societal impacts.

\bibliography{bias}

\begin{thebibliography}{61}
\expandafter\ifx\csname natexlab\endcsname\relax\def\natexlab#1{#1}\fi

\bibitem[{Abdollahpouri et~al.(2020)Abdollahpouri, Mansoury, Burke, and
  Mobasher}]{connection}
Himan Abdollahpouri, Masoud Mansoury, Robin Burke, and Bamshad Mobasher. 2020.
\newblock The connection between popularity bias, calibration, and fairness in
  recommendation.
\newblock In \emph{Proceedings of the 14th ACM Conference on Recommender
  Systems}, pages 726--731.

\bibitem[{Amig{\'o} et~al.(2023)Amig{\'o}, Deldjoo, Mizzaro, and
  Bellog{\'\i}n}]{unifying}
Enrique Amig{\'o}, Yashar Deldjoo, Stefano Mizzaro, and Alejandro
  Bellog{\'\i}n. 2023.
\newblock A unifying and general account of fairness measurement in recommender
  systems.
\newblock \emph{Information Processing \& Management}, 60(1):103115.

\bibitem[{Bobadilla et~al.(2013)Bobadilla, Ortega, Hernando, and
  Guti{\'e}rrez}]{RSsurvey}
Jes{\'u}s Bobadilla, Fernando Ortega, Antonio Hernando, and Abraham
  Guti{\'e}rrez. 2013.
\newblock Recommender systems survey.
\newblock \emph{Knowledge-based systems}, 46:109--132.

\bibitem[{Brown et~al.(2020)Brown, Mann, Ryder, Subbiah, Kaplan, Dhariwal,
  Neelakantan, Shyam, Sastry, Askell et~al.}]{gpt-3}
Tom Brown, Benjamin Mann, Nick Ryder, Melanie Subbiah, Jared~D Kaplan, Prafulla
  Dhariwal, Arvind Neelakantan, Pranav Shyam, Girish Sastry, Amanda Askell,
  et~al. 2020.
\newblock Language models are few-shot learners.
\newblock \emph{Advances in neural information processing systems},
  33:1877--1901.

\bibitem[{Bubeck et~al.(2023)Bubeck, Chandrasekaran, Eldan, Gehrke, Horvitz,
  Kamar, Lee, Lee, Li, Lundberg et~al.}]{gpt4}
S{\'e}bastien Bubeck, Varun Chandrasekaran, Ronen Eldan, Johannes Gehrke, Eric
  Horvitz, Ece Kamar, Peter Lee, Yin~Tat Lee, Yuanzhi Li, Scott Lundberg,
  et~al. 2023.
\newblock Sparks of artificial general intelligence: Early experiments with
  gpt-4.
\newblock \emph{arXiv preprint arXiv:2303.12712}.

\bibitem[{Chakraborty et~al.(2018)Chakraborty, Alam, Dey, Chattopadhyay, and
  Mukhopadhyay}]{adversarial2018}
Anirban Chakraborty, Manaar Alam, Vishal Dey, Anupam Chattopadhyay, and Debdeep
  Mukhopadhyay. 2018.
\newblock Adversarial attacks and defences: A survey.
\newblock \emph{arXiv preprint arXiv:1810.00069}.

\bibitem[{Chen et~al.(2023)Chen, Liu, Huang, Wu, Liu, Jiang, Pu, Lei, Chen,
  Wang et~al.}]{chen2023large}
Jin Chen, Zheng Liu, Xu~Huang, Chenwang Wu, Qi~Liu, Gangwei Jiang, Yuanhao Pu,
  Yuxuan Lei, Xiaolong Chen, Xingmei Wang, et~al. 2023.
\newblock When large language models meet personalization: Perspectives of
  challenges and opportunities.
\newblock \emph{arXiv:2307.16376}.

\bibitem[{Cheng et~al.(2016)Cheng, Koc, Harmsen, Shaked, Chandra, Aradhye,
  Anderson, Corrado, Chai, Ispir et~al.}]{deepmodel}
Heng-Tze Cheng, Levent Koc, Jeremiah Harmsen, Tal Shaked, Tushar Chandra,
  Hrishi Aradhye, Glen Anderson, Greg Corrado, Wei Chai, Mustafa Ispir, et~al.
  2016.
\newblock Wide \& deep learning for recommender systems.
\newblock In \emph{Proceedings of the 1st workshop on deep learning for
  recommender systems}, pages 7--10.

\bibitem[{De~Cao et~al.(2021)De~Cao, Izacard, Riedel, and Petroni}]{genre}
Nicola De~Cao, Gautier Izacard, Sebastian Riedel, and Fabio Petroni. 2021.
\newblock Autoregressive entity retrieval.
\newblock \emph{International Conference on Learning Representations (ICLR)
  2021}.

\bibitem[{Deldjoo et~al.(2021)Deldjoo, Anelli, Zamani, Bellogin, and
  Di~Noia}]{flexiblefairness}
Yashar Deldjoo, Vito~Walter Anelli, Hamed Zamani, Alejandro Bellogin, and
  Tommaso Di~Noia. 2021.
\newblock A flexible framework for evaluating user and item fairness in
  recommender systems.
\newblock \emph{User Modeling and User-Adapted Interaction}, pages 1--55.

\bibitem[{Devlin et~al.(2018)Devlin, Chang, Lee, and Toutanova}]{bert}
Jacob Devlin, Ming-Wei Chang, Kenton Lee, and Kristina Toutanova. 2018.
\newblock Bert: Pre-training of deep bidirectional transformers for language
  understanding.
\newblock \emph{arXiv preprint arXiv:1810.04805}.

\bibitem[{Dong et~al.(2020)Dong, Zhu, Cheng, Feng, Cai, He, Xu, and
  Wen}]{counterfactuallearning}
Zhenhua Dong, Hong Zhu, Pengxiang Cheng, Xinhua Feng, Guohao Cai, Xiuqiang He,
  Jun Xu, and Jirong Wen. 2020.
\newblock Counterfactual learning for recommender system.
\newblock In \emph{Fourteenth ACM Conference on Recommender Systems}, pages
  568--569.

\bibitem[{Ekstrand et~al.(2019{\natexlab{a}})Ekstrand, Burke, and
  Diaz}]{fairnessrecretrieval}
Michael~D Ekstrand, Robin Burke, and Fernando Diaz. 2019{\natexlab{a}}.
\newblock Fairness and discrimination in recommendation and retrieval.
\newblock In \emph{Proceedings of the 13th ACM Conference on Recommender
  Systems}, pages 576--577.

\bibitem[{Ekstrand et~al.(2019{\natexlab{b}})Ekstrand, Burke, and
  Diaz}]{ekstrand2019fairness}
Michael~D Ekstrand, Robin Burke, and Fernando Diaz. 2019{\natexlab{b}}.
\newblock Fairness and discrimination in retrieval and recommendation.
\newblock In \emph{Proceedings of the 42nd International ACM SIGIR Conference
  on Research and Development in Information Retrieval}, pages 1403--1404.

\bibitem[{Fan et~al.(2023)Fan, Zhao, Li, Liu, Mei, Wang, Tang, and
  Li}]{fan2023recommender}
Wenqi Fan, Zihuai Zhao, Jiatong Li, Yunqing Liu, Xiaowei Mei, Yiqi Wang,
  Jiliang Tang, and Qing Li. 2023.
\newblock Recommender systems in the era of large language models (llms).
\newblock \emph{arXiv:2307.02046}.

\bibitem[{Ganguli et~al.(2022)Ganguli, Lovitt, Kernion, Askell, Bai, Kadavath,
  Mann, Perez, Schiefer, Ndousse et~al.}]{ganguli2022red}
Deep Ganguli, Liane Lovitt, Jackson Kernion, Amanda Askell, Yuntao Bai, Saurav
  Kadavath, Ben Mann, Ethan Perez, Nicholas Schiefer, Kamal Ndousse, et~al.
  2022.
\newblock Red teaming language models to reduce harms: Methods, scaling
  behaviors, and lessons learned.
\newblock \emph{arXiv preprint arXiv:2209.07858}.

\bibitem[{Ge et~al.(2021)Ge, Liu, Gao, Xian, Li, Zhao, Pei, Sun, Ge, Ou
  et~al.}]{longtermfairness}
Yingqiang Ge, Shuchang Liu, Ruoyuan Gao, Yikun Xian, Yunqi Li, Xiangyu Zhao,
  Changhua Pei, Fei Sun, Junfeng Ge, Wenwu Ou, et~al. 2021.
\newblock Towards long-term fairness in recommendation.
\newblock In \emph{Proceedings of the 14th ACM international conference on web
  search and data mining}, pages 445--453.

\bibitem[{Ge et~al.(2022)Ge, Tan, Zhu, Xia, Luo, Liu, Fu, Geng, Li, and
  Zhang}]{explainable}
Yingqiang Ge, Juntao Tan, Yan Zhu, Yinglong Xia, Jiebo Luo, Shuchang Liu,
  Zuohui Fu, Shijie Geng, Zelong Li, and Yongfeng Zhang. 2022.
\newblock Explainable fairness in recommendation.
\newblock \emph{arXiv preprint arXiv:2204.11159}.

\bibitem[{Geng et~al.(2022)Geng, Liu, Fu, Ge, and Zhang}]{P5}
Shijie Geng, Shuchang Liu, Zuohui Fu, Yingqiang Ge, and Yongfeng Zhang. 2022.
\newblock Recommendation as language processing (rlp): A unified pretrain,
  personalized prompt \& predict paradigm (p5).
\newblock In \emph{Proceedings of the Sixteenth ACM Conference on Recommender
  Systems}.

\bibitem[{Geng and Liu(2023)}]{openlm2023openllama}
Xinyang Geng and Hao Liu. 2023.
\newblock \href {https://github.com/openlm-research/open_llama} {Openllama: An
  open reproduction of llama}.

\bibitem[{Harper and Konstan(2015)}]{ml}
F~Maxwell Harper and Joseph~A Konstan. 2015.
\newblock The movielens datasets: History and context.
\newblock \emph{Acm transactions on interactive intelligent systems (tiis)},
  5(4):1--19.

\bibitem[{Hidasi and Karatzoglou(2018)}]{rnntopk}
Bal{\'a}zs Hidasi and Alexandros Karatzoglou. 2018.
\newblock Recurrent neural networks with top-k gains for session-based
  recommendations.
\newblock In \emph{Proceedings of the 27th ACM international conference on
  information and knowledge management}, pages 843--852.

\bibitem[{Hidasi et~al.(2015)Hidasi, Karatzoglou, Baltrunas, and
  Tikk}]{session}
Bal{\'a}zs Hidasi, Alexandros Karatzoglou, Linas Baltrunas, and Domonkos Tikk.
  2015.
\newblock Session-based recommendations with recurrent neural networks.
\newblock \emph{arXiv preprint arXiv:1511.06939}.

\bibitem[{Hua et~al.(2023)Hua, Xu, Ge, and Zhang}]{hua2023index}
Wenyue Hua, Shuyuan Xu, Yingqiang Ge, and Yongfeng Zhang. 2023.
\newblock {How to Index Item IDs for Recommendation Foundation Models}.
\newblock In \emph{Proceedings of the 1st International ACM SIGIR Conference on
  Information Retrieval in the Asia Pacific (SIGIR-AP)}.

\bibitem[{Kang and McAuley(2018)}]{sasrec}
Wang-Cheng Kang and Julian McAuley. 2018.
\newblock Self-attentive sequential recommendation.
\newblock In \emph{2018 IEEE international conference on data mining (ICDM)},
  pages 197--206. IEEE.

\bibitem[{Koren et~al.(2009)Koren, Bell, and Volinsky}]{koren2009matrix}
Yehuda Koren, Robert Bell, and Chris Volinsky. 2009.
\newblock Matrix factorization techniques for recommender systems.
\newblock \emph{Computer}, 42(8):30--37.

\bibitem[{Leonhardt et~al.(2018)Leonhardt, Anand, and Khosla}]{userfairness}
Jurek Leonhardt, Avishek Anand, and Megha Khosla. 2018.
\newblock User fairness in recommender systems.
\newblock In \emph{Companion Proceedings of the The Web Conference 2018}, pages
  101--102.

\bibitem[{Lester et~al.(2021)Lester, Al-Rfou, and Constant}]{powerofscale}
Brian Lester, Rami Al-Rfou, and Noah Constant. 2021.
\newblock The power of scale for parameter-efficient prompt tuning.
\newblock \emph{EMNLP}.

\bibitem[{Li et~al.(2023{\natexlab{a}})Li, Zhang, Liu, and Chen}]{li2023large}
Lei Li, Yongfeng Zhang, Dugang Liu, and Li~Chen. 2023{\natexlab{a}}.
\newblock Large language models for generative recommendation: A survey and
  visionary discussions.
\newblock \emph{arXiv:2309.01157}.

\bibitem[{Li and Liang(2021)}]{prefixprompt}
Xiang~Lisa Li and Percy Liang. 2021.
\newblock Prefix-tuning: Optimizing continuous prompts for generation.
\newblock \emph{arXiv preprint arXiv:2101.00190}.

\bibitem[{Li et~al.(2023{\natexlab{b}})Li, Chen, Xu, Ge, Tan, Liu, and
  Zhang}]{li2023fairness}
Yunqi Li, Hanxiong Chen, Shuyuan Xu, Yingqiang Ge, Juntao Tan, Shuchang Liu,
  and Yongfeng Zhang. 2023{\natexlab{b}}.
\newblock Fairness in recommendation: Foundations, methods, and applications.
\newblock \emph{ACM Transactions on Intelligent Systems and Technology},
  14(5):1--48.

\bibitem[{Li et~al.(2021)Li, Chen, Xu, Ge, and Zhang}]{personalized}
Yunqi Li, Hanxiong Chen, Shuyuan Xu, Yingqiang Ge, and Yongfeng Zhang. 2021.
\newblock Towards personalized fairness based on causal notion.
\newblock In \emph{Proceedings of the 44th International ACM SIGIR Conference
  on Research and Development in Information Retrieval}, pages 1054--1063.

\bibitem[{Li and Zhang(2023)}]{li2023fairnesschatgpt}
Yunqi Li and Yongfeng Zhang. 2023.
\newblock Fairness of chatgpt.
\newblock \emph{arXiv:2305.18569}.

\bibitem[{Liang et~al.(2018)Liang, Zheng, Chen, Sangaiah, and Zhao}]{BiasedMF}
Nan Liang, Hai-Tao Zheng, Jin-Yuan Chen, Arun~Kumar Sangaiah, and Cong-Zhi
  Zhao. 2018.
\newblock Trsdl: Tag-aware recommender system based on deep
  learning--intelligent computing systems.
\newblock \emph{Applied Sciences}, 8(5):799.

\bibitem[{Liang et~al.(2022)Liang, Bommasani, Lee, Tsipras, Soylu, Yasunaga,
  Zhang, Narayanan, Wu, Kumar et~al.}]{liang2022holistic}
Percy Liang, Rishi Bommasani, Tony Lee, Dimitris Tsipras, Dilara Soylu,
  Michihiro Yasunaga, Yian Zhang, Deepak Narayanan, Yuhuai Wu, Ananya Kumar,
  et~al. 2022.
\newblock Holistic evaluation of language models.
\newblock \emph{arXiv preprint arXiv:2211.09110}.

\bibitem[{Lin et~al.(2023)Lin, Dai, Xi, Liu, Chen, Li, Zhu, Guo, Yu, Tang
  et~al.}]{lin2023can}
Jianghao Lin, Xinyi Dai, Yunjia Xi, Weiwen Liu, Bo~Chen, Xiangyang Li, Chenxu
  Zhu, Huifeng Guo, Yong Yu, Ruiming Tang, et~al. 2023.
\newblock How can recommender systems benefit from large language models: A
  survey.
\newblock \emph{arXiv:2306.05817}.

\bibitem[{Liu et~al.(2023)Liu, Zhang, and Gulla}]{liu2023pre}
Peng Liu, Lemei Zhang, and Jon~Atle Gulla. 2023.
\newblock Pre-train, prompt and recommendation: A comprehensive survey of
  language modelling paradigm adaptations in recommender systems.
\newblock \emph{arXiv:2302.03735}.

\bibitem[{Lowd and Meek(2005)}]{adversarial2005}
Daniel Lowd and Christopher Meek. 2005.
\newblock Adversarial learning.
\newblock In \emph{Proceedings of the eleventh ACM SIGKDD international
  conference on Knowledge discovery in data mining}, pages 641--647.

\bibitem[{Mao et~al.(2021)Mao, Zhu, Wang, Dai, Dong, Xiao, and He}]{simplex}
Kelong Mao, Jieming Zhu, Jinpeng Wang, Quanyu Dai, Zhenhua Dong, Xi~Xiao, and
  Xiuqiang He. 2021.
\newblock Simplex: A simple and strong baseline for collaborative filtering.
\newblock In \emph{Proceedings of the 30th ACM International Conference on
  Information \& Knowledge Management}, pages 1243--1252.

\bibitem[{Menon and Williamson(2018)}]{PMF}
Aditya~Krishna Menon and Robert~C Williamson. 2018.
\newblock The cost of fairness in binary classification.
\newblock In \emph{Conference on Fairness, Accountability and Transparency},
  pages 107--118. PMLR.

\bibitem[{Mnih and Salakhutdinov(2007)}]{mnih2007probabilistic}
Andriy Mnih and Russ~R Salakhutdinov. 2007.
\newblock Probabilistic matrix factorization.
\newblock \emph{Advances in neural information processing systems}, 20.

\bibitem[{Raffel et~al.(2020)Raffel, Shazeer, Roberts, Lee, Narang, Matena,
  Zhou, Li, Liu et~al.}]{t5}
Colin Raffel, Noam Shazeer, Adam Roberts, Katherine Lee, Sharan Narang, Michael
  Matena, Yanqi Zhou, Wei Li, Peter~J Liu, et~al. 2020.
\newblock Exploring the limits of transfer learning with a unified text-to-text
  transformer.
\newblock \emph{J. Mach. Learn. Res.}, 21(140):1--67.

\bibitem[{Rahmani et~al.(2022)Rahmani, Naghiaei, Dehghan, and
  Aliannejadi}]{userexperiments}
Hossein~A Rahmani, Mohammadmehdi Naghiaei, Mahdi Dehghan, and Mohammad
  Aliannejadi. 2022.
\newblock Experiments on generalizability of user-oriented fairness in
  recommender systems.
\newblock In \emph{Proceedings of the 45th International ACM SIGIR Conference
  on Research and Development in Information Retrieval}, pages 2755--2764.

\bibitem[{Ravfogel et~al.(2020)Ravfogel, Elazar, Gonen, Twiton, and
  Goldberg}]{nullitout}
Shauli Ravfogel, Yanai Elazar, Hila Gonen, Michael Twiton, and Yoav Goldberg.
  2020.
\newblock Null it out: Guarding protected attributes by iterative nullspace
  projection.
\newblock \emph{arXiv preprint arXiv:2004.07667}.

\bibitem[{Shrestha and Yang(2019)}]{algorithmicfairness}
Yash~Raj Shrestha and Yongjie Yang. 2019.
\newblock Fairness in algorithmic decision-making: Applications in multi-winner
  voting, machine learning, and recommender systems.
\newblock \emph{Algorithms}, 12(9):199.

\bibitem[{Sonboli et~al.(2021)Sonboli, Smith, Cabral~Berenfus, Burke, and
  Fiesler}]{userfairnesstransparency}
Nasim Sonboli, Jessie~J Smith, Florencia Cabral~Berenfus, Robin Burke, and
  Casey Fiesler. 2021.
\newblock Fairness and transparency in recommendation: The users’
  perspective.
\newblock In \emph{Proceedings of the 29th ACM Conference on User Modeling,
  Adaptation and Personalization}, pages 274--279.

\bibitem[{Sun et~al.(2019)Sun, Liu, Wu, Pei, Lin, Ou, and Jiang}]{bert4rec}
Fei Sun, Jun Liu, Jian Wu, Changhua Pei, Xiao Lin, Wenwu Ou, and Peng Jiang.
  2019.
\newblock Bert4rec: Sequential recommendation with bidirectional encoder
  representations from transformer.
\newblock In \emph{Proceedings of the 28th ACM international conference on
  information and knowledge management}, pages 1441--1450.

\bibitem[{Sun et~al.(2023)Sun, Zhang, Deng, Cheng, and Huang}]{sun2023safety}
Hao Sun, Zhexin Zhang, Jiawen Deng, Jiale Cheng, and Minlie Huang. 2023.
\newblock Safety assessment of chinese large language models.
\newblock \emph{arXiv preprint arXiv:2304.10436}.

\bibitem[{Touvron et~al.(2023)Touvron, Lavril, Izacard, Martinet, Lachaux,
  Lacroix, Rozi{\`e}re, Goyal, Hambro, Azhar et~al.}]{llama}
Hugo Touvron, Thibaut Lavril, Gautier Izacard, Xavier Martinet, Marie-Anne
  Lachaux, Timoth{\'e}e Lacroix, Baptiste Rozi{\`e}re, Naman Goyal, Eric
  Hambro, Faisal Azhar, et~al. 2023.
\newblock Llama: Open and efficient foundation language models.
\newblock \emph{arXiv preprint arXiv:2302.13971}.

\bibitem[{Wang and Yu(2019)}]{deeplearningadversarial}
Huaxia Wang and Chun-Nam Yu. 2019.
\newblock A direct approach to robust deep learning using adversarial networks.
\newblock \emph{arXiv preprint arXiv:1905.09591}.

\bibitem[{Wang et~al.(2023)Wang, Ma, Zhang, Liu, and Ma}]{wang2023survey}
Yifan Wang, Weizhi Ma, Min Zhang, Yiqun Liu, and Shaoping Ma. 2023.
\newblock A survey on the fairness of recommender systems.
\newblock \emph{ACM Transactions on Information Systems}, 41(3):1--43.

\bibitem[{Wu et~al.(2017)Wu, Ahmed, Beutel, Smola, and Jing}]{rrn}
Chao-Yuan Wu, Amr Ahmed, Alex Beutel, Alexander~J Smola, and How Jing. 2017.
\newblock Recurrent recommender networks.
\newblock In \emph{Proceedings of the tenth ACM international conference on web
  search and data mining}, pages 495--503.

\bibitem[{Wu et~al.(2021)Wu, Cao, Xu, and Tan}]{twoside}
Yao Wu, Jian Cao, Guandong Xu, and Yudong Tan. 2021.
\newblock Tfrom: A two-sided fairness-aware recommendation model for both
  customers and providers.
\newblock In \emph{Proceedings of the 44th International ACM SIGIR Conference
  on Research and Development in Information Retrieval}, pages 1013--1022.

\bibitem[{Wu et~al.(2022)Wu, Xie, Zhu, Zhuang, Xiang, Zhang, Lin, and
  He}]{selective}
Yiqing Wu, Ruobing Xie, Yongchun Zhu, Fuzhen Zhuang, Ao~Xiang, Xu~Zhang, Leyu
  Lin, and Qing He. 2022.
\newblock Selective fairness in recommendation via prompts.
\newblock In \emph{Proceedings of the 45th International ACM SIGIR Conference
  on Research and Development in Information Retrieval}, pages 2657--2662.

\bibitem[{Xu et~al.(2023)Xu, Hua, and Zhang}]{xu2023openp5}
Shuyuan Xu, Wenyue Hua, and Yongfeng Zhang. 2023.
\newblock Openp5: Benchmarking foundation models for recommendation.
\newblock \emph{arXiv preprint arXiv:2306.11134}.

\bibitem[{Yi et~al.(2019)Yi, Shen, Liu, Zhang, Zhang, Liu, and Xiong}]{DMF}
Baolin Yi, Xiaoxuan Shen, Hai Liu, Zhaoli Zhang, Wei Zhang, Sannyuya Liu, and
  Naixue Xiong. 2019.
\newblock Deep matrix factorization with implicit feedback embedding for
  recommendation system.
\newblock \emph{IEEE Transactions on Industrial Informatics}, 15(8):4591--4601.

\bibitem[{Yu et~al.(2016)Yu, Liu, Wu, Wang, and Tan}]{dynamic}
Feng Yu, Qiang Liu, Shu Wu, Liang Wang, and Tieniu Tan. 2016.
\newblock A dynamic recurrent model for next basket recommendation.
\newblock In \emph{Proceedings of the 39th International ACM SIGIR conference
  on Research and Development in Information Retrieval}, pages 729--732.

\bibitem[{Zhang et~al.(2023)Zhang, Bao, Zhang, Wang, Feng, and
  He}]{zhang2023chatgpt}
Jizhi Zhang, Keqin Bao, Yang Zhang, Wenjie Wang, Fuli Feng, and Xiangnan He.
  2023.
\newblock Is chatgpt fair for recommendation? evaluating fairness in large
  language model recommendation.
\newblock \emph{RecSys}.

\bibitem[{Zhang et~al.(2017)Zhang, Ai, Chen, and Croft}]{zhang2017joint}
Yongfeng Zhang, Qingyao Ai, Xu~Chen, and W~Bruce Croft. 2017.
\newblock Joint representation learning for top-n recommendation with
  heterogeneous information sources.
\newblock In \emph{Proceedings of the 2017 ACM on Conference on Information and
  Knowledge Management}, pages 1449--1458.

\bibitem[{Zhao et~al.(2022)Zhao, Alwidian, and Mahmoud}]{adversarial2022}
Weimin Zhao, Sanaa Alwidian, and Qusay~H Mahmoud. 2022.
\newblock Adversarial training methods for deep learning: A systematic review.
\newblock \emph{Algorithms}, 15(8):283.

\bibitem[{Zhuo et~al.(2023)Zhuo, Huang, Chen, and Xing}]{zhuo2023exploring}
Terry~Yue Zhuo, Yujin Huang, Chunyang Chen, and Zhenchang Xing. 2023.
\newblock Exploring ai ethics of chatgpt: A diagnostic analysis.
\newblock \emph{arXiv preprint arXiv:2301.12867}.

\end{thebibliography}

\newpage
\appendix

\section*{APPENDIX}

\section{Probing Unfairness in LLM-based RS}
\label{appendix:probing}
Probing the user attributes out of LLM is a nontrivial task in LLM-based RS because each user does not have one specific user embedding. In this section, we illustrate three methods to detect unfairness of LLM-based RS.
The results show that even if the training data does not explicitly use user-sensitive attributes, LLM-based RS still implicitly infers user information and possibly leaks it. 

In general, there are three distinct methodologies for probing user attributes in LLM: (1) eliciting attributes through in-context learning utilizing interpretable discrete prompts that are manually designed, (2) eliciting attributes through the training of tunable prompts, and in this paper, we adopt soft prompts which are more amenable to optimization compared with discrete prompts, (3) training a classifier on embeddings generated for user tokens that appear in the input prompts. The three subsections below show how much user attribute information is encoded and how they can be probed by the three methods above. 


\subsection{Manually-Designed Prompt}
In the first method, we directly adopt manually-designed discrete prompts using in-context learning to probe user sensitive attributes out of the LLM. We use questions about users with (or without) their item interaction history and expect reasonable answers when multiple examples are appended in the input. 

More specifically, we test two types of manual prompts: direct prompts and in-context learning prompts. The direct prompt directly asks the LLM about a user's sensitive attribute, as shown by the following example, one without user-item interaction and one with user-item interaction.

\begin{formal}
\small
\textbf{\textit{Discrete Prompt without User-Item Interaction}}\\
\textbf{Input}: What is the \{\{attribute\}\} for user\_\{\{user\_id\}\}? \textbf{Output}: \{\{user attribute value\}\}\\
\textbf{\textit{Discrete Prompt with User-Item Interaction}}\\
\textbf{Input}: User\_\{\{user\_id\}\} has watched movies (or bought insurance) \{\{sequence of movie (or insurance) IDs\}\}. What is the \{\{attribute\}\} of user\_\{\{user\_id\}\}? \textbf{Output}: \{\{user attribute value\}\}
\end{formal}

The attribute can be gender, age, occupation or marital status provided by MovieLens and Insurance datasets. The answer template is simply the value of the questioned attribute, such as female / male, above / below 55 years old, or single / married. 
We constrain the output generated from the decoder based on constrained token generation over all possible values of the questioned attribute \cite{genre}.

\begin{figure}[t]
    \centering
    \includegraphics[width=0.45\textwidth]{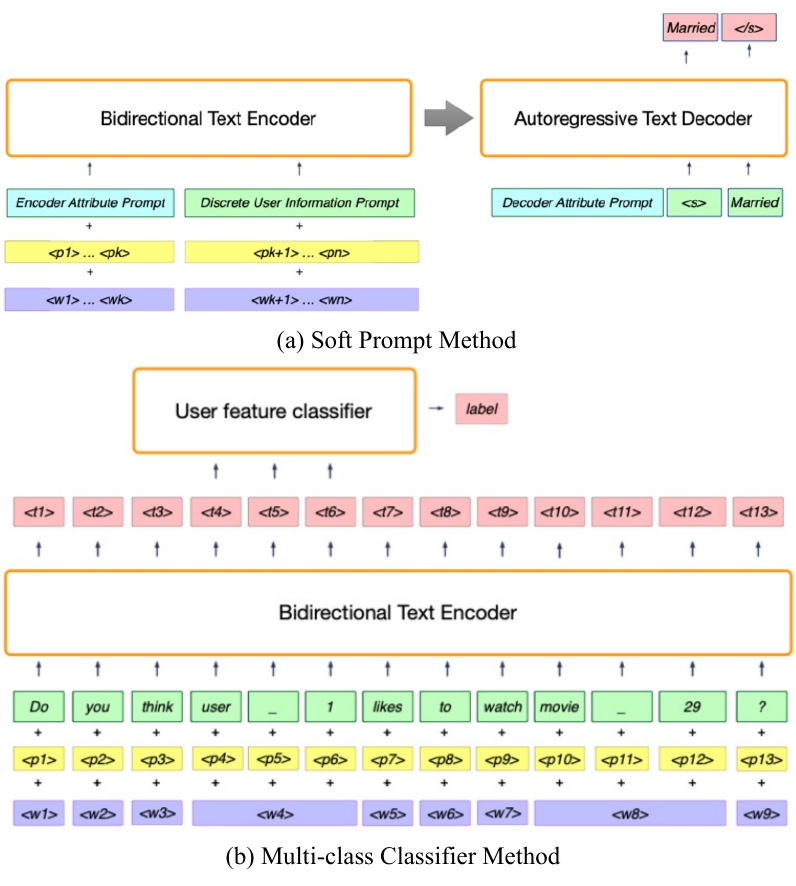}
    \vspace{-10pt}
    \caption{Details for Probing Methods}
    \label{fig:probing_method}
\vspace{-15pt}
\end{figure}

For in-context learning prompts, contextual examples, which are question-answer pairs of randomly sampled known users, are appended before the question. We use as many contextual examples as the maximum input length allows.
The following example presents in-context learning prompts for the MovieLens dataset with and without user-item interaction information. We use gray color to differentiate the context from the question.

\begin{formal}
\small
\textbf{\textit{In-context Learning Example w/o User-Item Interaction}}\\
\textbf{Input}: \textcolor{gray}{What is the gender of user\_1? Female. What is the gender of user\_2? Male. What is the gender of user\_3? Female. What is the gender of user\_4? Female. What is the gender of user\_5? Male.} What is the gender of user\_10? \textbf{Output}: Male\\
\textbf{\textit{In-context Learning Example w/ User-Item Interaction}}\\
\textbf{Input}: \textcolor{gray}{User\_1 has watched movies 17, 1991, 29, 3039, 890. What is the gender of user\_1? Female. User\_2 has watched movies 29, 1084, 27, 93, 781. What is the gender of user\_2? Male.} User\_10 has watched movies 136, 798, 2778, 1894, 1. What is the gender of user\_10? \textbf{Output}: Male.
\end{formal}

We measure the performance of probing user sensitive attributes from LLM using AUC and results are presented in Table \ref{tab:manual}.
We notice that the AUC is either 50\% or slightly above 50\%, indicating that the prediction result is no better than random guessing. Thus even if there is user sensitive information encoded in LLM such as P5 (see the next two subsections), direct prompting cannot elicit it. The reason may be that the model is trained using numerical user and item identifiers rather than natural language labels or descriptions and does not include any additional user or item metadata. Therefore, prompts designed using natural language may not align with the numerical representations used in the model's training.
Manual prompts' failure can be considered as an advantage of LLM-based RS, as user attributes will not be leaked too easily. 

\begin{table}[t]
\small
    \centering
    \setlength{\tabcolsep}{4pt}
    \begin{tabular}{l|c|c|c|c}
    \toprule
         MovieLens & Gender & Age & Occupation & --\\
        \hline
        w/ interaction & 50.33 & 50.09 & 50.00 & --\\
        w/o interaction & 50.26 & 50.00 & 50.00 & --\\
        \hline
        \hline
        Insurance & Gender & Age & Occupation & Marital \\
        \hline
         w/ interaction & 50.00 & 50.33 & 50.47 & 50.20 \\
        w/o interaction & 50.00 & 50.00 & 50.00 & 50.00 \\
         \bottomrule
    \end{tabular}
    \vspace{-10pt}
    \caption{Manually-Designed Prompt AUC (\%)}
    \label{tab:manual}
    \vspace{-10pt}
\end{table}

\subsection{Soft Probing Prompt Tuning}
In the second method, we adopt tunable prompts proposed in \citeauthor{powerofscale} to explore soft prompt tuning with a frozen pre-trained LLM-based RS to elicit attributes. Each attribute has one soft probing prompt trained, which is tailored to act as a question, guiding the model to produce desired outcomes. Soft probing prompts can be optimized end-to-end over a training dataset and can condense information by learning from the training. The model structure is presented in Figure \ref{fig:probing_method}(a). 
The encoder input is a concatenation of an encoder attribute prompt and an untunable discrete prompt, where the discrete prompt part includes the target user and relevant user-item interaction history, as shown below:
\begin{formal}
\small
User user\_\{\{user\_id\}\} has watched movies (or bought insurances) \{\{sequence of item IDs\}\}.
\end{formal}
The decoder attends to the decoder attribute prompt, the previously generated tokens, and the encoder hidden state to predict the probability distribution of future tokens. The encoder attribute prompt and decoder attribute prompt are generated respectively by a two-layer multi-layer perceptron (MLP) and a three-layer MLP as proposed in \cite{prefixprompt}. The prompts are tuned by minimizing the negative
log-likelihood of the attribute value tokens $y$ conditioned on the input text $x$ and the soft probing prompts $p$ in an end-to-end manner:
\begin{equation}
    L = -\sum\nolimits_{j=1}^{|y|}\log P(y_j|y_{<j}, x, p)
\end{equation}

For answer generation, we also apply the constrained generation as in manual prompting. 

In experiments, we create separate train and test datasets by dividing all users into two groups in a 9:1 ratio, and generating a unique discrete attribute prompt for each user in the process.
Experimental results on MovieLens and Insurance datasets are shown in Table \ref{tab:soft_auc}. We notice that using soft probing prompt tuning does generate non-trivial predictions on user attributes, especially on MovieLens dataset, indicating that LLM-based RS does encode user attributes and leaks personal information.

\begin{table}[t]
\small
    \centering
    \setlength{\tabcolsep}{4pt}
    \begin{tabular}{l|c|c|c|c}
    \toprule
         \multirow{2}{*}{MovieLens} & gender & age & occupation & --\\
         \cline{2-5}
         & 70.84 & 64.60 & 56.50 & --\\
        \hline 
        \hline 
        \multirow{2}{*}{Insurance} & gender & age & occupation & marital \\
        \cline{2-5}
         & 50.00 & 51.80 & 50.00 & 70.28 \\
         \bottomrule
    \end{tabular}
    \vspace{-10pt}
    \caption{Soft Probing Prompt Tuning AUC (\%)}
    \label{tab:soft_auc}
\end{table}

\begin{table}[t]
\small
    \centering
    \setlength{\tabcolsep}{4pt}
    \begin{tabular}{l|c|c|c|c}
    \toprule
         \multirow{2}{*}{MovieLens} & gender & age & occupation & -- \\
        \cline{2-5}
        & 74.71 & 67.40 & 53.47 & --\\
        \hline 
        \hline 
        \multirow{2}{*}{Insurance} & gender & age & occupation & marital \\
        \cline{2-5}
        & 50.13 & 56.92 & 57.87 & 76.37 \\
         \bottomrule
    \end{tabular}
    \vspace{-10pt}
    \caption{Multi-class Classifier AUC (\%)}
    \label{tab:classifier_auc}
    \vspace{-10pt}
\end{table}

\subsection{Multi-Class Classifier}
The third probing method trains a multi-class classifier on the user token embeddings generated by the encoder for all input sentences in the training set.
The model structure is presented in Figure \ref{fig:probing_method}(b), where the classifier is a seven-layer multi-layer perceptron (MLP) network trained by standard cross-entropy loss. 
Tables \ref{tab:classifier_auc} presents the AUC results.
The non-trivial AUC scores indicate that LLM-based RS also suffers from user information leakage, similar to other RS models. We also observe that the AUC scores obtained from the trained classifier tend to be higher than those obtained through soft probing prompt tuning.
This suggests that training a classifier is a more effective probing method of user sensitive attributes from LLMs than training soft probing prompts. This observation highlights that the cross-entropy loss over multiple classes is better suitable than the negative log-likelihood loss over the entire vocabulary. 
This observation is leveraged in our design of fairness-aware foundation model architecture.

\subsection{\mbox{Summary of Probing LLM-RS Unfairness}}
This section demonstrates three possible methods to elicit user sensitive attributes from LLM-based RS: manually-designed discrete prompts, soft probing prompts, and multi-class classifier.
The latter two successfully generate non-trivial user attribute values among the three methods. Figure \ref{fig:auc} illustrates the degree of unfairness on LLM models trained on MovieLens and Insurance datasets, measured by the AUC of label prediction. The model on MovieLens is unfair on gender, age, and slightly on occupation, while the model on Insurance is unfair on the marital status the most.


\section{Results on P5-OpenLlama-3B}
\label{llama_result}
This appendix presents all the experiment results of the P5 recommendation paradigm under the Open-llama-3B backbone. The observations here are largely consistent with that under the T5 backbone.

Table \ref{tab:llama-matching} and Table \ref{tab:llama-sequential} present the recommendation performance and AUC scores.
\begin{table}[ht]
    \centering
    \small
    \setlength{\tabcolsep}{1.5pt}
    \begin{tabular}{l|c|c}
    \toprule
    \textbf{Dataset} & \textbf{MovieLens} & \textbf{Insurance}\\
        \hline 
        $\uparrow$ Hit@1 & 22.79 & 83.01\\
        $\uparrow$ Hit@3 & 35.97 & 87.95\\
        $\uparrow$ Hit@10 & 62.18 & 87.95\\
        \hline
        $\downarrow$ AUC (G) & 73.39 & 50.49\\
        $\downarrow$ AUC (A) & 59.59 & 51.68\\
        $\downarrow$ AUC (O) & 50.43 & 50.18\\
        $\downarrow$ AUC (M) & -- & 58.40\\
        \bottomrule
    \end{tabular}
    \caption{Results of matching-based recommendation, G means Gender, A means Age, O means Occupation, and M means Marital Status (\%).}
    \label{tab:llama-matching}
\end{table}

\begin{table}[ht]
    \centering
    \small
    \setlength{\tabcolsep}{1.5pt}
    \begin{tabular}{l|c|c}
    \toprule
    \textbf{Dataset} & \textbf{MovieLens} & \textbf{Insurance}\\
        \hline 
        $\uparrow$ Hit@1 & 33.70 & 84.17\\
        $\uparrow$ Hit@3 & 46.92 & 87.23\\
        $\uparrow$ Hit@10 & 68.18 & 90.11\\
        \hline
        $\downarrow$ AUC (G) & 73.39 & 51.32\\
        $\downarrow$ AUC (A) & 59.59 & 52.40\\
        $\downarrow$ AUC (O) & 50.43 & 50.97\\
        $\downarrow$ AUC (M) & -- & 61.89\\
        \bottomrule
    \end{tabular}
    \caption{Results of sequential-based recommendation, G means Gender, A means Age, O means Occupation, and M means Marital Status (\%).}
    \label{tab:llama-sequential}
\end{table}

Tables \ref{tab:llama_debias_matching} and \ref{tab:llama_debias_sequential} present the single-attribute fairness performance using single-attribute CFPs.

\begin{table}[!ht]
\small
    \centering
    \setlength{\tabcolsep}{1pt}
    \resizebox{\linewidth}{!}{
    \begin{tabular}{l|c|c|c|c|c|c}
    \toprule
        \textbf{Dataset} & \multicolumn{3}{c|}{\textbf{MovieLens}} & \multicolumn{3}{c}{\textbf{Insurance}}\\
        \hline 
        \textbf{Attribute} & \textbf{Gender} & \textbf{Age} & \textbf{Occupation} & \textbf{Age} & \textbf{Marital} & \textbf{Occupation} \\
        \hline
        $\uparrow$ Hit@1 & 20.78 & 22.08 & 22.79 & 83.01 & 82.74 & 83.01 \\
        $\uparrow$ Hit@3 & 34.62 & 35.12 & 35.97 & 87.95 & 87.31 & 87.95 \\
        $\uparrow$ Hit@10 & 59.14 & 60.97 & 62.18 & 87.95 & 87.92 & 87.95\\
        $\downarrow$ AUC & 52.30 & 50.23 & 50.43 & 51.68 & 50.00 & 50.18\\
        \bottomrule
    \end{tabular}
    }
    \caption{Results of single-attribute fairness-aware prompting on matching-based models (\%)}
    \label{tab:llama_debias_matching}
\end{table}

\begin{table}[!ht]
\small
    \centering
    \setlength{\tabcolsep}{1pt}
    \resizebox{\linewidth}{!}{
    \begin{tabular}{l|c|c|c|c|c|c}
    \toprule
        \textbf{Dataset} & \multicolumn{3}{c|}{\textbf{MovieLens}} & \multicolumn{3}{c}{\textbf{Insurance}}\\
        \hline 
        \textbf{Attribute} & \textbf{Gender} & \textbf{Age} & \textbf{Occupation} & \textbf{Age} & \textbf{Marital} & \textbf{Occupation} \\
        \hline
        $\uparrow$ Hit@1 & 31.72 & 32.69 & 33.70 & 84.17 & 82.33 & 84.17\\
        $\uparrow$ Hit@3 & 44.60 & 45.72 & 46.92 & 87.23 & 86.14 & 87.23\\
        $\uparrow$ Hit@10 & 65.13 & 67.73 & 68.18 & 90.11 & 88.90 & 90.11\\
        $\downarrow$ AUC & 54.38 & 52.25 & 50.43 & 52.40 & 50.23 & 50.97\\
        \bottomrule
    \end{tabular}
    }
    \caption{Results of single-attribute fairness-aware prompting on sequential models (\%)}
    \label{tab:llama_debias_sequential}
\end{table}

\begin{table}[!ht]
\small
    \centering
    \setlength{\tabcolsep}{5pt}
    \begin{tabular}{l|c|c|c|c}
    \toprule
    \textbf{Model} & \textbf{GA} & \textbf{GO} & \textbf{AO} & \textbf{GAO} \\
    \hline 
        $\uparrow$ Hit@1 & 22.13 & 20.78 & 22.08 & 22.13\\
        $\uparrow$ Hit@3 & 36.77 & 34.62 & 35.12 & 36.77\\
        $\uparrow$ Hit@10 & 60.08 & 59.14 & 60.97 & 60.08\\
        $\downarrow$ Avg. AUC & 50.49 & 51.37 & 50.33 & 50.47 \\
        \bottomrule
    \end{tabular}
    \caption{Results of multi-attribute fairness-aware prompting on MovieLens dataset (\%)}
    \label{tab:llama_combine_debias_matching_ml}
\end{table}

\begin{table}[!ht]
\small
    \centering
    \setlength{\tabcolsep}{5pt}
    \begin{tabular}{l|c|c|c|c}
    \toprule
    \textbf{Model} & \textbf{AO} & \textbf{AM} & \textbf{MO} & \textbf{AMO} \\
    \hline 
        $\uparrow$ Hit@1 & 84.17 & 82.33 & 82.33 & 82.33 \\
        $\uparrow$ Hit@3 & 87.23 & 86.14 & 86.14 & 86.14\\
        $\uparrow$ Hit@10 & 90.11 & 88.90 & 88.90 & 88.90\\
        $\downarrow$ Avg. AUC & 51.69 & 51.32 & 50.60 & 51.20\\
        \bottomrule
    \end{tabular}
    \caption{Results of multi-attribute fairness-aware prompting on Insurance dataset (\%)}
    \label{tab:llama_combine_debias_matching_insurance}
\end{table}

Tables \ref{tab:llama_combine_debias_matching_ml} and \ref{tab:llama_combine_debias_matching_insurance} present the multi-attribute fairness-aware performance using prompt mixture over multiple CFPs.

\section{Pseudo Code for CFP Training}
\label{algorithm}
In this section, we provide the pseudo code of training the Counterfactually-Fair Prompts (CFP) for unbiased recommendation foundation model.

\begin{algorithm}[H]
\small
\caption{CFP Training
}\label{alg:cap}
\begin{algorithmic}[1]
\Require Pretrained LLM4RS $\mathcal{M}$, Randomly initialized prefix prompt $\mathcal{P}$, Randomly initialized classifier $\mathcal{C}$, discriminator loss weight $\lambda$, number of epochs $Epoch\_num$, number of steps $T$ to update $\mathcal{C}$ on $L_{dis}$ or prefix prompt $\mathcal{P}$ on $L_{rec}$, number of batches $R$ to update prefix prompt $\mathcal{P}$ on adversarial loss $L$

\For{epoch $\gets$ 1\,to\,$Epoch\_num$}

    \For{batch\_num, batch}

                \For{i $\in$ $[1, T]$}    
                
                    \State rec\_loss, u\_emb $\gets$ $\mathcal{P}$($\mathcal{M}$,\text{batch})
                
                    \State dis\_loss $\gets$ $\mathcal{C}$(\text{u\_emb}, \text{label}\_u)
                
                    \State $L$ $\gets$ rec\_loss - $\lambda$ $\cdot$ dis\_loss
                        
                    \State Optimize $\mathcal{P}$ based on $L$ with $\mathcal{M}, \mathcal{C}$ fixed
                    
                    \EndFor

                \If{batch\_num \% $R$ == 0}
                
                    \For{i $\in$ [1, T]} 
                    
                        \State rec\_loss $\gets$ $\mathcal{P}$($\mathcal{M}$,\text{batch})
                            
                        \State Optimize $\mathcal{P}$  on rec\_loss with $\mathcal{M}, \mathcal{C}$ fixed
                    
                        \EndFor

                    \For{i $\in$ [1, T]} 
                
                        \State rec\_loss, u\_emb $\gets$ $\mathcal{P}$($\mathcal{M}$,\text{batch})    
                
                        \State dis\_loss $\gets$ $\mathcal{C}$(\text{u\_emb}, \text{label}\_u)
                    
                        \State Optimize $\mathcal{C}$  on dis\_loss with $\mathcal{M}, \mathcal{P}$ fixed
                    
                        \EndFor
                
                \EndIf
\EndFor
\EndFor    
\end{algorithmic}
\end{algorithm}

\end{document}